\documentclass[lettersize,journal]{IEEEtran}
\IEEEoverridecommandlockouts
\usepackage{cite}
\usepackage{adjustbox}
\usepackage{amsmath,amssymb,amsfonts}
\usepackage{algorithmicx}
\usepackage{algorithm}
\usepackage{algpseudocode}
\usepackage{graphicx}
\usepackage{textcomp}
\usepackage{paralist}
\usepackage{algorithmicx}
 \usepackage{listings}
\usepackage{xspace}
\usepackage{paralist}
\usepackage{blindtext}
\usepackage{multicol}
\usepackage{algpseudocode}
\usepackage{graphicx}
\usepackage{amssymb}
\usepackage{mathtools}
\usepackage{amsmath}
\usepackage{amsfonts}
\usepackage{titlesec}
\usepackage{gensymb}
\usepackage{indentfirst}
\usepackage{float}
\usepackage{booktabs, multirow}
\usepackage{soul}
\usepackage[table]{xcolor}
\usepackage{changepage,threeparttable}
\usepackage{xcolor}
\usepackage{array}
\usepackage{graphicx}
\usepackage{subcaption}
\usepackage{booktabs, multirow} 
\usepackage{soul}
\usepackage{changepage,threeparttable}
\usepackage{bm}
\usepackage{paralist}
\usepackage{blindtext}
\usepackage{multicol}
\usepackage{graphicx}
\newtheorem{thm}{Theorem}[section]
\algrenewcommand\alglinenumber[1]{\footnotesize #1:}

\def\BibTeX{{\rm B\kern-.05em{\sc i\kern-.025em b}\kern-.08em
    T\kern-.1667em\lower.7ex\hbox{E}\kern-.125emX}}
\begin{document}

\title{Bayesian framework for characterizing cryptocurrency market dynamics, structural dependency, and volatility using potential field}

\author{Anoop C V, Neeraj Negi, Anup Aprem
\IEEEcompsocitemizethanks{\IEEEcompsocthanksitem The authors are with the Department of Electronics and Communication Engineering, National Institute of Technology Calicut,
India.\protect\\
E-mail: $\{$anoop\_p210065ec,neeraj\_m210444ec,anup.aprem$\}$$@$nitc.ac.in
}}
\maketitle
\begin{abstract}
Identifying the structural dependence between the cryptocurrencies and predicting market trend are fundamental for effective portfolio management in cryptocurrency trading. 
  In this paper, we present a unified Bayesian framework based on potential field theory and Gaussian Process to characterize the structural dependency of various cryptocurrencies, using historic price information. 
The following are our significant contributions:
\begin{inparaenum}[(i)]
    \item Proposed a novel model for cryptocurrency price movements as a trajectory of a dynamical system governed by a time-varying non-linear potential field.
    \item Validated the existence of the non-linear potential function in cryptocurrency market through Lyapunov stability analysis.
    \item Developed a Bayesian framework for inferring the non-linear potential function from observed cryptocurrency prices.
    \item Proposed that attractors and repellers inferred from the potential field are reliable cryptocurrency market indicators, surpassing existing attributes, such as, mean, open price or close price of an observation window, in the literature.
    \item Analysis of cryptocurrency market during various Bitcoin crash durations from April~$2017$ to November~$2021$, shows that attractors captured the market trend, volatility, and correlation. In addition, attractors aids explainability and visualization.
    \item The structural dependence inferred by the proposed approach was found to be consistent with results obtained using the popular wavelet coherence approach.
    \item The proposed market indicators (attractors and repellers) can be used to improve the prediction performance of state-of-art deep learning price prediction models. As, an example, we show improvement in Litecoin price prediction up to a horizon of $12$ days.
\end{inparaenum}
\end{abstract}
\begin{IEEEkeywords}
Cryptocurrency market analysis, Structural dependency, Bayesian Data Analysis, Potential Field Method, Gaussian Process, Uncertainty Characterization.
\end{IEEEkeywords}
\section{Introduction}
Cryptocurrency trading, which has gained enormous popularity in recent years, is viewed as a high-risk, high-return investment because of volatility, wherein price fluctuates considerably even within an hour. The volatility of cryptocurrency prices is attributed to many factors such as unregulated nature of the market, sensitivity to market perception and news, fixed availability of cryptocurrency, and the shift in investor sentiment driven by speculative trading activities by peers, among others \cite{CHEAH2015volReasons}.
The easiness, of converting from one cryptocurrency to another, offered by cryptocurrency exchanges, and the flexibility for fractional ownership, wherein the investors are allowed to buy and sell cryptocurrencies in fractional amounts, facilitating the investment of even small sum of money, are notable advantages of cryptocurrency trading over traditional stock market trading.
The crashes of cryptocurrencies, characterized by a cryptocurrency's price dropping by over 15$\%$ within a period of 24 hours or longer, are significant events for cryptocurrency traders \cite{CrashDef}. By identifying structural interdependence between various cryptocurrencies and price change patterns during these crashes, investors can effectively diversify their portfolios, a well-known strategy in volatile and uncertain markets. In addition, the structural interdependence between various cryptocurrencies have shown significant changes before and after the occurrences of crash of major cryptocurrencies such as Bitcoin \cite{CrashDef}. 

In addition, experienced and novice investors prefer to use Automated Financial Consulting (AFC) options for cryptocurrency trading, often called as \emph{trading bots} or \emph{algorithmic trading}, that can automate cryptocurrency portfolio management~\cite{fang2022cryptTradingSurv}. 
In the year $2022$, trading bots accounted for more than $60\%$ of all cryptocurrency trading activity \cite{cryptobots2023proportion}. 
Trading bots, typically, use the following strategies \cite{fang2022cryptTradingSurv}: 
\begin{inparaenum}[(i)]
    \item continuously track cryptocurrency price movements and dynamically buy, sell, or hold the cryptocurrency to make profits by taking advantage of the volatility of cryptocurrencies,
    \item operate in accordance with the hedge investment strategy, maintaining a diversified portfolio by investing in non-correlated cryptocurrencies, to lower the risk associated with an unanticipated crash in price.
\end{inparaenum}These trading bots are implemented using statistical and machine learning (ML) algorithms for risk assessment and portfolio management.

Hence, the requirements of cryptocurrency trading include: \begin{inparaenum}[(i)]
    \item the ability to recognize structural dependencies or correlations between individual cryptocurrencies, to do effective portfolio management.
    \item the capability of predicting price patterns (including both rises and falls).
\end{inparaenum}
However, practical realization of the aforementioned requirements are challenging due to a multitude of reasons, including the following: 
the performance of parametric price prediction models based on price history is poor (accuracy of Machine Learning based models in predicting cryptocurrency price is less than $65\%$) \cite{sanders2017garbage,ML12021}. Also, since, a large fraction of cryptocurrency market investors are novices in cryptocurrency trading, market psychology plays a predominant role in the dynamics of cryptocurrency market, wherein the trading activities of the investors are driven by the market sentiments. Hence, approaches, that uses traditional market fundamentals such as supply, demand, cost of production and number of competitors, fail to model the cryptocurrency price evolution accurately. 

In this context, the following are the significant contributions of this paper:
\begin{enumerate}
    \item Proposed a novel model where the price movements of the cryptocurrencies are governed by a time-varying non-linear potential field function. The potential field theory treats the price of various cryptocurrencies as a trajectory of a chaotic dynamical system. 
    \item Validated the existence of a non-linear potential function through Lyapunov stability analysis of observed cryptocurrency price trajectory. 
    \item Developed a Bayesian framework for inferring the non-linear potential function from the observed cryptocurrency prices.  
    \item The framework of potential field theory lends itself to the notion of attractors and repellers, the states, towards which a dynamical system  will eventually converge, and, away from which the time series will diverge, respectively. We propose that attractors and repellers are more reliable cryptocurrency market indicators than existing attributes in the literature such as  open, close, or mean prices.
    \item Analysis of the cryptocurrency market data during different Bitcoin crash durations, from April $2017$ to November $2021$, showed that attractors or repellers captured the market trend and volatility. In addition, such a characterization aids explainability and  visualization through: \begin{inparaenum}[(i)]
        \item inferring the direction of cryptocurrency price correction and the corresponding confidence level,
        \item characterizing the volatility and correlation of cryptocurrency prices,
        \item inferring the direction of asset movement in cryptocurrency market,
        \item analysis of convergence characteristics of the market,
        \item predicting the temporal evolution of the market trends.
    \end{inparaenum}
    \item The proposed market indicators (attractors and repellers of the inferred potential field) yield better performance when used as a feature for price prediction in existing deep learning models. As an example,  using it  as an input feature to the hybrid LSTM model in \cite{existingLSTM} gives better performance for Litecoin prediction up to a horizon of $12$ days.
    \item The structural dependence, inferred using the proposed approach, is compared with the results obtained using the widely used wavelet coherence approach in \cite{existingWavelet,CrashDef} and results obtained were found to be consistent. However, unlike wavelet coherence analysis, the proposed approach can be readily generalized for more than $2$ cryptocurrencies, a disadvantage of the wavelet coherence approach.  
\end{enumerate}

The paper is structured as follows: Section \ref{LitRev} provides an overview of the relevant literature. The theoretical aspects of the proposed methodology are discussed in Section \ref{PropMeth}, which is divided into several subsections. Section \ref{LyaTheory}, proposes the non-linear potential function model, and, establishes the existence of a valid potential function, using Lyapunov stability analysis. Section \ref{PF} defines attractors and repellers of the potential field using the canonical vector calculus operations. Section \ref{LapGP} presents the theoretical aspects of the Bayesian framework for inferring the potential field. Section \ref{KL} elaborates on the probabilistic approach for inferring attractors and repellers. The numerical results are presented in detail in Section \ref{Result}. Subsection \ref{dataset} describes the cryptocurrency datasets and the various Bitcoin crash durations, that this paper focuses on. Subsection \ref{Validation} presents the numerical results and inferences drawn from Lyapunov stability analysis. Subsection \ref{2D_analysis} explains the numerical results and insights of cryptocurrency market analysis of two cryptocurrencies (Bitcoin and Ethereum) during the various Bitcoin crash durations. Subsection \ref{3D_analysis} extends the market analysis results to top 10 cryptocurrencies. Subsection \ref{WaveletCompar} examines the consistency between the proposed potential field approach and the results obtained from the wavelet coherence approach in characterizing structural dependencies. Subsection \ref{LSTMComp} discusses the performance improvement in price prediction of existing deep learning models when the inferred mean attractor is used as one of the input features. Finally Section~\ref{Concl} offers concluding remarks. 

\section{Literature Review}\label{LitRev}
\subsection{Cryptocurrency vs Stock market}
The majority of the investors in cryptocurrency market are lay investors, with none or limited prior experience in cryptocurrency trading. The phenomena of market psychology and peer influence are more predominant in cryptocurrency market than conventional stock markets~\cite{CrashDef}. Given the significant differences between the conventional stock market and the cryptocurrency market, the tools, and approaches that were developed and deployed for the stock market cannot be readily applied for the cryptocurrency market~\cite{all_model_survey,PriceDriverscryptocurrency}. However, cryptocurrency trading bots \cite{2022SentiTradingBotML,2023AutoMLTrading,2023RLTrading}, are preferred by novice and experienced traders to automate the cryptocurrency exchange and portfolio management. Inferring the structural dependence between the cryptocurrencies, and predicting their price movement, are two important aspects that facilitate efficient portfolio management \cite{lee2017allFactors}.

\subsection{Correlation and Structural Dependency of Cryptocurrencies}
A plethora of literature exists on the topic of structural dependence between various cryptocurrencies, with wavelet coherence approach~\cite{CrashDef,existingWavelet,PriceDriverscryptocurrency}, emerging as the widely used statistical approach, because of its ability to handle non stationarity of cryptocurrency market as well as in capturing both long term and short term dependencies. Generalized Autoregressive Conditional Heteroscedasticity (GARCH) \cite{antonakakis2019GARCH,gao2021hybrid} \cite{correl2023WCGARCH}, which is mainly used to infer the volatility of returns, is also capable of characterizing structural dependence between cryptocurrencies~\cite{correl2023WCGARCH}. Time Lagged Cross Correlation (TLCC)~\cite{TLCC} and entropy based measures~\cite{entropyCorrel} are some alternative methods for finding correlation between cryptocurrencies. However, the correlation analysis using all these approaches, have a common limitation of having to consider distinct pairwise combinations of cryptocurrencies, which is not a scalable approach as the number of cryptocurrencies is high as $1800$ (as of March $2022$) \cite{alatorre2023stocks}.
\subsection{Prediction of Cryptocurrency Prices:}
Another line of work uses statistical, and ML based techniques for prediction of price movements or volatility in cryptocurrency market. The two commonly used metrics used to evaluate performance of price prediction models are: Mean Absolute Error (MAE) and Mean Squared Error (MSE). MAE captures the average performance of a price predictive model~\cite{willmott2005advantagesMAE}, whereas, MSE represents the ability of the model in capturing extreme events such as crash~\cite{chai2014RMSEAdv}. 

Compared to classical machine learning models such as linear regression, quadratic discriminant analysis and support vector machines, recurrent neural networks were found to be performing better, in terms of MSE and MAE, because of its ability to capture long term dependencies~\cite{EconometricsVsLSTM,mudassir2020time,sezer2020financial}. Similarly, RNN based techniques were found to perform better compared to statistical econometric models such as GARCH. Within RNN based techniques, Long Short Term Memory (LSTM) \cite{existingLSTM,all_model_survey} is the most popular model for cryptocurrency price prediction -- refer to recent survey in~\cite{murray2023StatMLDLEns} for a comparative study between various models. However, the error performance of models, that predict the cryptocurrency price, solely based on historic prices, has enormous scope for improvement, especially in capturing extreme events like cryptocurrency crashes~\cite{murray2023StatMLDLEns}.

\subsection{Input Features of the Cryptocurrency Price Prediction Models}Various models, in literature, available for cryptocurrency price prediction, differs from one another mainly in terms of the input features used. The factors that influence the cryptocurrency price can be either internal or external factors~\cite{all_model_survey}. Internal factors include technology related aspects of cryptocurrency, such as, coin circulation, reward system and mining difficulty, and are mostly deterministic. The external factors include various sub-factors, that reflect the market demand and investor sentiment, such as popularity of a cryptocurrency, speculations about market, gold price, restrictions imposed by government, legalizations, Google Trend~\cite{hassani2018GoogleTrend} and, volume and sentiments of tweets in Twitter~\cite{abraham2018Twitter,critien2022Twitter2}, among others. Since market psychology and peer influence play an important role in the dynamics of cryptocurrency market, the factors driving cryptocurrency price change from time to time, making it challenging to construct a reliable and sustainable predictive model.
In their respective works, \cite{lee2017allFactors} and \cite{all_model_survey} evaluated the performance of various cryptocurrency price prediction models based on the choice of input features. They concluded that incorporating a combination of time series of historical prices and other internal and external factors that influence cryptocurrency prices as input features can enhance the performance of predictive models, compared to models based solely on any one of these features. 

\subsection{Use of Structural Dependence Information in Cryptocurrency Price Prediction Models:}There are recent attempts to supplement the ML models with structural dependency model to further improve the accuracy of cryptocurrency price prediction. \cite{zhong2023lstmSD} uses the correlation metric obtained using Graph Attention Network (GAT) to serve as an additional input feature to an LSTM based predictive model, while, \cite{existingLSTM}, presented a hybrid approach, that exploits the structural dependence between LTC and BTC, to further improve the performance of LSTM based approaches, in LTC price prediction. \cite{existingLSTM} used time series of opening and mean prices of hourly windows to make predictions over, $1$, $3$, $7$ and $30$ days. In addition to the historic prices of LTC (referred as child coin), the direction of price movement of BTC (referred as parent coin) within 1-hour windows was also used as an input feature to the model in \cite{existingLSTM}. However, the opening and mean price need not necessarily represent the exact market trend or dynamics and structural dependence during the observation window, since the cryptocurrency prices are highly volatile, and hence may affect the reliability of the predictions made. Further, given a finite length time series of cryptocurrency price observation, the question, `\emph{which attributes of the cryptocurrency time series will reliably represent the market dynamics and the volatility characteristics?}', is still an open research problem.

This motivates the need for a unified framework, that can reliably quantify the market trend and underlying uncertainty, while characterizing the structural dependence between various cryptocurrencies, given any time series of cryptocurrency price observations.  

\section{Proposed Methodology: The Potential Field Approach}\label{PropMeth}
The proposed approach, in this paper, models the dynamics of the cryptocurrency market using a time-varying potential function over an $M$-dimensional state space of prices. 
The cryptocurrency price, at each time, can be considered to be analogous to the position of a hypothetical particle of unit mass on a space-time surface, governed by a gravitational potential field. A particle in a potential field tends to move towards it's minima, defined as attractors or sink, and away from it's maxima, defined as repeller or source. 

Sec.~{\ref{LyaTheory}} defines the potential field model, considered in this paper, and establishes of existence of a potential field in the cryptocurrency market, through Lyapunov stability analysis from non-linear dynamical systems. In Sec.~\ref{LapGP}, we consider the problem of inference of the potential field using Bayesian machine learning, specifically using Gaussian processes. In Sec.~\ref{KL}, we consider the problem of identifying significant attractors and repellers of the potential field,  which we postulate as characterizing the structural dependency of the cryptocurrency market and providing information about the market trend.

\subsection{Potential field model for the cryptocurrency market: Lyapunov stability analysis}\label{LyaTheory}
The time evolution of a hypothetical unit mass, in general, can be described by the following set of nonlinear differential equations 
\begin{equation}\label{fx}
    \dot{\mathbf{x}} = \frac{d\mathbf{x}}{dt} = f(\mathbf{x},t),
\end{equation}
where, $\mathbf{x}\in \mathbb{R}^M$ is a state space vector, and the vector function $f(\mathbf{\cdot})$ determines the system's evolution over time. 
A potential field, if it exists, $\phi(\cdot)$, is a function whose gradient, $\nabla\phi(\cdot)$, gives the force acting on the particle, where, $\nabla$, is the spatial derivative operator. A valid potential function, $\phi(t, \mathbf{x})$, if it exists, is related to the vector function $f(\mathbf{x},t)$ as
\begin{equation}\label{fxphirel}
    -\nabla\phi(t,\mathbf{x}) = \frac{df\left(\mathbf{x},t\right)}{dt} = \frac{d^2\mathbf{x}}{dt^2},
\end{equation}
i.e.\  the negative of the spatial derivative (gradient) of a potential function is the same as the time derivative of $f(.)$ at a given time. 

Let $M$ and $N$ denote the total number of cryptocurrencies and the total number of price observations, respectively. At a given time\footnote{In this paper, we assume that time is discretized, with some sampling interval $\Delta t$.} $t_n$, $n = 1, 2, \cdot\cdot\cdot, N$, the vector of prices of all cryptocurrencies (position vector of unit mass) can be denoted as $\mathbf{x}_n = [x_{1,n}, x_{2,n}, \cdot\cdot\cdot, x_{M,n}]^T$. The sequential solution of \eqref{fx}, starting from an initial state, $\mathbf{x}_1$, forms a trajectory, denoted by $\boldsymbol{\zeta}_{1:N} = \{\mathbf{x}_1,\cdot\cdot\cdot,\mathbf{x}_N\}$. 

The existence of a non-linear potential function, as in \eqref{fxphirel} is given by the following theorem \cite{LyapunovProof}: 
\begin{thm}\label{Theorem1}
    \textit{For a certain dynamical system, \begin{inparaenum}[(i)]
        \item explicit formulation of a function of the form \eqref{fx} can be given for any global Lyapunov function of the system,
        \item any potential function $\phi$ obtained from \eqref{fxphirel} is a global Lyapunov function, and
        \item if a trajectory belongs to a stable dynamic system in the \emph{Lyapunov sense}, it is equivalent to the existence of a non-linear potential function, with a stable state, defined in its state space.
    \end{inparaenum}}
\end{thm}
The proof of Theorem \ref{Theorem1}, available in \cite{LyapunovProof} and \cite{ma2014potential}, is derived from the notion of conservation of total energy, of the system given by \eqref{fx}. 

A system is said to be \emph{stable} in the Lyapunov sense, if the separation between two trajectories of the system with nearby\footnote{Two states in a state space are said to be nearby, if the separation between them is less than a small positive quantity, $\epsilon$.} initial states converge to a single point, namely the attractor. If two nearby trajectories diverge away as time evolves, the system is said to be \emph{unstable}. Lyapunov Stability analysis \cite{vulpiani2009chaos,LyapExpData,LyapExp2022} is a technique from chaos theory \cite{chaostext} to determine the stability of the non-linear dynamic system in~\eqref{fx}. Let $\delta_{\mathbf{x}}(t)$ be the difference vector at time $t$ between two nearby trajectories that start at $\mathbf{x}_i$ and $\mathbf{x}_i + \delta_{\mathbf{x}}(i)$, respectively. The average exponential rate of growth or decay of the norm, $|\delta_{\mathbf{x}}(t)|$ over time, is given by:

\begin{equation}
    |\delta_{\mathbf{x}}(t)| = \exp\left(\lambda t\right)|\delta_{\mathbf{x}}(i)|,
\end{equation}
where, $\lambda$ is the Lyapunov exponent, and is defined as
\begin{equation}\label{Lexp}
    \lambda = \lim_{t \to \infty}\left[\frac{1}{t}\right]\log\frac{|\delta_{\mathbf{x}}(t)|}{|\delta_{\mathbf{x}}(i)|}, 
\end{equation}
i.e., we can compute the Lyapunov exponents by comparing the time evolution of a perturbed trajectory with the original trajectory, that starts at some point $\mathbf{x}_i$. 

If we do not have the functional form of $f(\mathbf{x},t)$ and only a sample trajectory is available, we can apply Algorithm \ref{AlgLexp}, described in \cite{LyapExp1993practical}, to estimate the Lyapunov exponents from the given trajectory. Algorithm \ref{AlgLexp} takes $\boldsymbol{\zeta}_{1:N}$ as input and generates a list of Lyapunov exponents, $L$. It proceeds by identifying all possible pairs of points $\mathbf{x}_i$ and $\mathbf{x}_j$, in the trajectory, such that $i<j$ and $i<(N-k)$, where $k$ is a small positive integer that denotes the sampling interval in terms of number of samples. Further, it determines the distance between the subsequent points that are integer multiples of $k$ time steps ahead, starting at all $\mathbf{x}_i$ and $\mathbf{x}_j$, in order to calculate the exponent estimates in equation \eqref{Lexp}. 

\begin{algorithm}[H]
{
\fontsize{9pt}{9pt}\selectfont
\caption{Finding Lyapunov Exponents from Trajectory \cite{LyapExp1993practical}}\label{AlgLexp}
\begin{algorithmic}[1]
\Require $\boldsymbol{\zeta}_{1:N} = \{\mathbf{x}_I\}_{i = 1}^N$, $\epsilon$, $k$, $\Delta t$
\State $i\gets1$,  $L\gets\{\}$
\While{$i \le N-k$}
    \State $j\gets i+ 1$
    \While{$j\le N$}
        \If{$||\mathbf{x}_i-\mathbf{x}_j||\le\epsilon$}
            \State $p\gets1$, $N_{ij}\gets\lfloor \left(N-j\right)/k\rfloor$, $\lambda_{ij}\gets0$
            \While{$p\le N_{ij}$}
                \State $\lambda_p\gets\left[\frac{1}{pk\Delta t}\right]\log\left[\frac{||\mathbf{x}_{i+kp} - \mathbf{x}_{j+kp}||}{||\mathbf{x}_i-\mathbf{x}_j||}\right]$
                \State $\lambda_{ij}\gets\lambda_{ij}+\lambda_p$
                \State $p\gets p+1$
            \EndWhile 
        \EndIf 
        \State $L\gets\{L,\frac{\lambda_{ij}}{N_{ij}}\}$, $j\gets j+1$
    \EndWhile
        \State $i\gets i+1$
\EndWhile
\State Output: $L$\label{LyaOut}
\end{algorithmic}}
\end{algorithm}

The largest Lyapunov exponent, $\lambda_1 = max(L)$, determines the rate of divergence or convergence of nearby trajectories in state space. If $\lambda_1$ is negative, then nearby trajectories converge towards the attractor\footnote{An attractor is a set of points in state space towards which a system evolves over time. More specifically, an attractor is a subset of the state space that is invariant under the dynamics of the system, meaning that once a trajectory enters the attractor, it stays there indefinitely or until it is perturbed by some external influence \cite{chaostext}.}. Conversely, if $\lambda_1$ is positive, then nearby trajectories diverge away, indicating the system is unstable. Since Lyapunov stability imply the existence of a valid potential function according to Theorem~\ref{Theorem1}, if the trajectory belongs to a Lyapunov stable system, the attractor inferred from the trajectory, will be a stable state towards which the market tends to stabilize during the observation window.

\subsection{Laplacian of a Scalar Potential Field and the Attractor}\label{PF}
This section defines the concepts of attractors and repellers of a potential field with the help of canonical vector calculus operations on the potential gradient. The potential gradient, $\nabla\phi$, in \eqref{fxphirel}, at $\mathbf{x} = [x_1, x_2,\cdot\cdot\cdot,x_N]^T$, at any given time, $t$, can be represented as a vector in \eqref{grad}
\begin{equation}\label{grad} 
    \nabla\phi = \left[\frac{\partial\phi}{\partial x_1},\frac{\partial\phi}{\partial x_2},\cdot\cdot\cdot,\frac{\partial\phi}{\partial x_M}\right]^T,
\end{equation}where, the time and state arguments of the potential functions are suppressed for the convenience of representation. 
The negative of the gradient vector, $-\nabla\phi$, is a vector field, normally visualized as lines of flux, represents the amount of acceleration and the direction of the evolution, of the state, under the influence of the potential field. 
The, divergence, of any vector field $\mathbf{F}$, denoted by $\nabla\cdot\mathbf{F}$, represents the density of a vector field's outward flux from an infinitesimally small volume surrounding a certain point. The Laplacian of a scalar potential function ($\phi$) is the divergence of the potential gradient, that quantifies the net out-flux from an infinitesimal hypercube centred at a point in the state space, and can be evaluated as in \eqref{LaplacianDef}:
\begin{equation}\label{LaplacianDef}
    \nabla^2\phi = \nabla.\nabla\phi = \sum_{i = 1}^{M}\frac{\partial^2\phi}{\partial x_i^2}.
\end{equation}

When the Laplacian of the potential field, at a particular point in the state space, is positive it signifies that the net out-flux, in that region, is positive. This, in turn, indicates that the net force exerted on the particle will be oriented outward, propelling the particle away from that region -- such regions are classified as repellers or sources. Conversely, if the Laplacian is negative, it denotes that the net force acting on a particle will be directed inward, leading to the stabilization of the particle in that region -- such regions are referred to as attractors or sinks. Hence, the attractors and repellers can be estimated as the regions in the state space with negative and positive Laplacian values, respectively. 

The Bayesian approach adopted for inferring the potential function and the attractors and repellers of the potential function from the observed trajectory is discussed in Sec.~{\ref{LapGP}}.
\subsection{Bayesian approach for inferring the potential function}\label{LapGP}
The trajectory~$\boldsymbol{\zeta}_{1:N}$ gives noisy observations of~$f$ as in~\eqref{fx}, which, in turn, can be used to estimate the potential gradient~$\nabla\phi$ using~\eqref{fxphirel}. 
Let, 
\begin{equation}\label{y}
    \mathbf{y}_i = \nabla\phi(\mathbf{x}_i) + \eta, \quad \eta \sim \mathcal{N}\left(0,\sigma^2\right),
\end{equation}
where, $\nabla\phi$ is the gradient of unknown potential function, which maps any state $\mathbf{x}$ to the corresponding potential gradient at $\mathbf{x}$, and the noise term $\eta$ characterize the randomness in the price observation as well as that incurred in the estimation of the gradient. 

We model, $\nabla\phi$, as a multi-input multi-output Gaussian Process (GP) \cite{GP,schulz2018GPtutorial}, which is a distribution over functions, as
\begin{equation}\label{GP}
    \nabla\phi(\mathbf{x}) \sim GP(\boldsymbol{\mu}(\mathbf{x}),\mathbf{k}(\mathbf{x},\mathbf{x}')),
\end{equation}
where, $\boldsymbol{\mu}(\mathbf{x}) = \mathbb{E}(\nabla\phi(\mathbf{x}))$, is the mean function, and $\mathbf{k}(\mathbf{x},\mathbf{x}')$ is the covariance function. In our work, $\boldsymbol{\mu}(\mathbf{x})$ is set to zero. The covariance function~$\mathbf{k}(\mathbf{x},\mathbf{x}')$, also known as the GP kernel, characterizes the similarity between the values of the function at any two distinct input states, $\mathbf{x}$ and $\mathbf{x}'$, and is defined as 
\begin{equation}\label{kernTheo}
    \mathbf{k}\left(\mathbf{x},\mathbf{x}'\right) = \mathbb{E}\left[\left(\nabla\phi(\mathbf{x})-\boldsymbol{\mu}(\mathbf{x})\right)^T\left(\nabla\phi(\mathbf{x}')-\boldsymbol{\mu}(\mathbf{x}')\right)\right].
\end{equation}
In practice, the function $\mathbf{k}$ is chosen based on the anticipated nature of the function to be inferred. In our context, we use a squared exponential (SE) kernel, under the assumption of a smooth infinitely differentiable potential function \cite{GPML}. The extension to other GP kernels is straightforward. The SE kernel is given by 
\begin{equation}\label{SEkernel}
  \hspace{-0.75ex}    \mathbf{k}_{SE}(\mathbf{x},\mathbf{x}') = \sigma_{SE}^2\exp{\left(-\frac{1}{2}\left(\mathbf{x}-\mathbf{x}'\right)^T\Lambda ^{-1}\left(\mathbf{x}-\mathbf{x}'\right)\right)},
\end{equation}
where, $\Lambda = \text{diag}(\lambda_1,\cdots,\lambda_M)\in\mathbb{R}^{M\times M}$ is a diagonal matrix of input scale length hyperparameters~$\{\lambda_i\}_{i=1}^M$ and $\sigma_{SE}$ is the signal variance hyperparameter. 

Let $\mathbf{X}_{\zeta} = [\mathbf{x}_1^T,\cdot\cdot\cdot,\mathbf{x}_N^T]^T$ be a matrix, with each row representing the transpose of the state vector in the trajectory, and $\mathbf{Y}_{\zeta} = [\mathbf{y}_1^T,\cdot\cdot\cdot,\mathbf{y}_N^T]^T$, the matrix of noisy observations of potential gradients. 
Given a set of $Q$ new states (test points) in the state space, represented as a matrix $\mathbf{X}^* = [\mathbf{x^*}_1^T,\cdot\cdot\cdot,\mathbf{x^*}_Q^T]^T$, we can infer the corresponding potential gradient by computing the posterior distribution, $p(\nabla\Phi(\mathbf{X^*})|\mathcal{D}_{\zeta})$. By definition of GP, $Y_{\zeta}\in\mathbb{R}^{N\times M}$ and $\nabla\phi(\mathbf{X^*})\in\mathbb{R}^{Q\times M}$ are jointly distributed as
\begin{equation}\label{posDis}
\hspace{-0.5em}
  \left[ {\begin{array}{c}
    \mathbf{Y}_{\zeta} \\
    \nabla\phi(\mathbf{x^*}) \\
  \end{array} } \right] = \mathcal{N}\left(0,\left[{\begin{array}{cc}
  \mathbf{K}(\mathbf{X_{\zeta}},\mathbf{X_{\zeta}}) + \sigma^2\mathbf{I} &\mathbf{K}(\mathbf{X_{\zeta}},\mathbf{X}^*)\\
  \mathbf{K}(\mathbf{X}^*,\mathbf{X_{\zeta}}) & \mathbf{K}(\mathbf{X}^*,\mathbf{X}^*)\\
  \end{array}}
  \right]\right),
\end{equation}
where, $\mathbf{K}(\mathbf{X_{\zeta}},\mathbf{X_{\zeta}})\in\mathbb{R}^{N\times N}$ is the covariance matrix between the training points, $\mathbf{K}(\mathbf{X_{\zeta}},\mathbf{X}^*)\in\mathbb{R}^{N\times Q}$ is the covariance matrix between all past observations and all test points, $\mathbf{K}(\mathbf{X}^*,\mathbf{X_{\zeta}})\in\mathbb{R}^{Q\times N}$ is the covariance matrix between all the test points and the training, $\mathbf{K}(\mathbf{X}^*,\mathbf{X}^*)\in\mathbb{R}^{Q\times Q}$ is the covariance matrix between all test points, $\sigma^2$ is the variance of $\eta$ and $\mathbf{I}\in\mathbb{R}^{N\times N}$ is an identity matrix. The posterior mean and variance of $\nabla\phi(\mathbf{X}^*)$ can be computed as \eqref{GPosMu} and \eqref{GPosVar}, respectively.
\begin{equation}\label{GPosMu}
\hspace{-1.01em}
    \mathbb{E}\left[\nabla\phi(\mathbf{X}^*)\right] = \mathbf{K}\left({\mathbf{X}_{\zeta},\mathbf{X}^{*}}\right)^{T}\left(\mathbf{K}\left({\mathbf{X}_{\zeta},\mathbf{X}_{\zeta}}\right) + \boldsymbol{\sigma}^{2}\mathbf{I}\right)^{-1}\mathbf{Y}_{\zeta},
\end{equation}
\begin{equation}\label{GPosVar}
    \begin{split}
        &V\left[\nabla\phi(\mathbf{X}^*)\right] = \mathbf{K}\left({\mathbf{X}_{\zeta},\mathbf{X}^{*}}\right) - \\&\mathbf{K}\left({\mathbf{X}_{\zeta},\mathbf{X}^{*}}\right)^{T}
        \left(\mathbf{K}\left({\mathbf{X}_{\zeta},\mathbf{X}_{\zeta}}\right) + \boldsymbol{\sigma}^{2}\mathbf{I}\right)^{-1}\mathbf{K}\left({\mathbf{X}_{\zeta},\mathbf{X}^{*}}\right).
    \end{split}
\end{equation}
Assuming $\nabla\phi$ at any given state is given by \eqref{grad}, its Jacobian matrix can be expressed as
\begin{equation}\label{Jacob}\begin{split}
J\left(\nabla\phi\right) &= \frac{\partial\nabla\phi}{\partial\mathbf{x^*}} \\&=
    \left[{\begin{array}{cccc}
       \frac{\partial^2\phi}{\partial x_1^2}  & \frac{\partial^2\phi}{\partial x_1\partial x_2} &\cdots & \frac{\partial^2\phi}{\partial x_1\partial x_M}\\
       \frac{\partial^2\phi}{\partial x_2\partial x_1}  & \frac{\partial^2\phi}{\partial x_2^2} &\cdots & \frac{\partial^2\phi}{\partial x_2\partial x_M}\\
       \vdots & \vdots & \vdots & \vdots \\
       \frac{\partial^2\phi}{\partial x_M\partial x_1}  & \frac{\partial^2\phi}{\partial x_M \partial x_2} &\cdots & \frac{\partial^2\phi}{\partial x_M^2}\\
    \end{array}}\right].
    \end{split}
\end{equation}
Since the derivative operation is an affine transformation, and an affine transformation of a GP is again a GP, the individual elements of the Jacobian Matrix in \eqref{Jacob}, at each test state $\mathbf{x^*}$, can be inferred from the derivative of the GP that models $\nabla\phi$. The squared exponential kernel for the first and second order derivatives, respectively, of the GP in \eqref{GP}, are given by \eqref{dSE} and \eqref{ddSE}.
\begin{equation}\label{dSE}
    \frac{\partial \mathbf{k}(\mathbf{x^*},\mathbf{x})}{\partial \mathbf{x^*}} = \Lambda^{-1}\left(\mathbf{x^*}-\mathbf{x}\right)\mathbf{k}(\mathbf{x^*},\mathbf{x})\in \mathbb{R}^M.
\end{equation}
\begin{equation}\label{ddSE}
    \frac{\partial^2\mathbf{k}(\mathbf{x^*},\mathbf{x})}{\partial\mathbf{x^*}\partial\mathbf{x^*}} 
    =\Lambda^{-1}\left((\mathbf{x^*}-\mathbf{x})(\mathbf{x^*}-\mathbf{x})^T\Lambda^{-1} - \mathbf{I}\right)\mathbf{k}(\mathbf{x^*},\mathbf{x}).
\end{equation}
In \eqref{dSE} and \eqref{ddSE}, $\mathbf{x^*}$ and $\mathbf{x}$ are test and train input states, respectively. The posterior mean, and the associated covariance of $J(\nabla\phi)$, at a test input $\mathbf{x^*}$, can be computed as \eqref{JacMu} and \eqref{JacCov}.
\begin{equation}\label{JacMu}
    \begin{split}   
        \mathbb{E}\left[\frac{\partial\nabla\phi(\mathbf{x}^*)}{\partial\mathbf{x}^*}\right] = \frac{\partial\mathbf{K}\left({\mathbf{X}_{\zeta},\mathbf{x}^*}\right)^{T}}{\partial\mathbf{x}^*}\left(\mathbf{K}\left({\mathbf{X}_{\zeta},\mathbf{X}_{\zeta}}\right) + \sigma^{2}\mathbf{I}\right)^{-1}\mathbf{Y}_{\zeta},
    \end{split}
\end{equation}
\begin{equation}\label{JacCov}
    \begin{split}
        &V\left[\frac{\partial\nabla\phi(\mathbf{x}^*)}{\partial\mathbf{x}^*}\right] = \frac{\partial^2\mathbf{K}({\mathbf{x}^*,\mathbf{x}^*})}{\partial{\mathbf{x}^*}\partial{\mathbf{x}^*}} - \\&\frac{\partial\mathbf{K}({\mathbf{X}_{\zeta},\mathbf{x}^*})^T}{\partial\mathbf{x}^*}
        \left(\mathbf{K}({\mathbf{X}_{\zeta},\mathbf{X}_{\zeta}}) + \sigma^{2}\mathbf{I}\right)^{-1}\frac{\partial\mathbf{K}({\mathbf{X}_{\zeta},\mathbf{x}^*})}{\partial\mathbf{x}^*},
    \end{split}
\end{equation}
where $\frac{\partial\mathbf{K}\left({\mathbf{X},\mathbf{x}^*}\right)}{\partial\mathbf{x}^*}$ in \eqref{JacMu}, is the derivative covariance matrix between all input training points, $\mathbf{X}_{\zeta}$ and the test point, $\mathbf{x^*}$, computed using the derivative kernel in \eqref{dSE}. Similarly, $\frac{\partial^2\mathbf{K}({\mathbf{x}^*,\mathbf{x}^*})}{\partial{\mathbf{x}^*}\partial{\mathbf{x}^*}}$ is the derivative covariance, between the test input $\mathbf{x^*}$ and itself, computed using \eqref{ddSE}. \eqref{JacMu} and \eqref{JacCov} gives the posterior inferences about all the elements in \eqref{Jacob}. Since, the Laplacian in \eqref{LaplacianDef} is the trace of the Jacobian matrix in \eqref{Jacob}, the posterior mean, $\boldsymbol{\mu}_{po}^{\nabla^2},$ and the variance, ${\sigma^2_{po}}^{\nabla^2},$ of the Laplacian, at the test point, $\mathbf{x}_i^* = [x_{i,1}^*,\cdots,x_{i,M}^*]^T$, is given by \eqref{LapMu} and \eqref{LapCov}, respectively.
\begin{equation}\label{LapMu}
    \boldsymbol{\mu}_{i,po}^{\nabla^2} = \mathbb{E}\left[\sum_{j=1}^M\frac{\partial^2\phi({x}_{i,j}^*)}{\partial x_j^2}\right].
\end{equation}
\begin{equation}\label{LapCov}
    \sigma_{i,po}^{\nabla^2} = Var\left[\sum_{j=1}^M\frac{\partial^2\phi({x}_{i,j}^*)}{\partial x_j^2}\right].
\end{equation}

In the next section, we deal with the problem of identifying attractors and repellers of the inferred potential function.
\subsection{Identifying Mean Attractors and Repellers of the Potential Function}\label{KL}

Even though it is possible to infer the potential field distribution in the entire continuous state space, it is not feasible to utilize the inferred potential field distribution directly as an input feature for standard machine learning models, which typically operate on finite-dimensional features. Hence, we propose using mean attractor, which can be inferred from potential field distribution, as a reliable feature capable of capturing the cryptocurrency market dynamics. This section will delve into the methodology for inferring the mean attractors and repellers of the potential field.

The prior distribution over the Laplacian of the potential field (before observing the data) at any $\mathbf{x}^* \in \mathbf{X}^*$ is a univariate normal distribution, with mean, $\boldsymbol{\mu}_{pr}^{\nabla^2} = 0$ and variance ${\sigma^2_{pr}}^{\nabla^2} = \sigma^2$. Given a set of observations $\{\mathbf{x}_n\}_{i=1}^N$, the GP model in \eqref{GP} can infer the posterior distributions of $\nabla^2\phi(t, \mathbf{x}^*)$, $\forall \mathbf{x}^* \in \mathbf{X}^*$, using \eqref{LapMu} and \eqref{LapCov}. The posterior distribution, at any $\mathbf{x}^* \in \mathbf{X}^*$, is also a univariate normal distribution with a mean, $\boldsymbol{\mu}_{po}^{\nabla^2}$, and variance, ${\sigma^2_{po}}^{\nabla^2}$ computed at $\mathbf{x}^*$. In this paper, we use Kullback-Leibler (KL) divergence to identify significant attractors, as in \cite{CobbPotential}. KL divergence quantifies the degree of dissimilarity between two distributions and the KL divergence between prior and posterior distributions of the Laplacian at any location can be computed as 
\begin{equation}\label{KLDeq}
    \begin{split}    
    &D_{KL}\left(p_{prior}||p_{posterior}\right)
    = \\&\frac{1}{2}\left(\frac{{\sigma^2_{pr}}^{\nabla^2}}{{\sigma^2_{po}}^{\nabla^2}} + \frac{(\boldsymbol{\mu}_{po}^{\nabla^2}-\boldsymbol{\mu}_{pr}^{\nabla^2})^2}{{\sigma^2_{po}}^{\nabla^2}} - 1 + \log\left(\frac{\sigma_{po}^{\nabla^2}}{\sigma_{pr}^{\nabla^2}}\right)\right).
    \end{split}
\end{equation}

The rationale for using KL divergence as a measure of significance for identifying attractors is the following. $\boldsymbol{\mu}_{pr}^{\nabla^2} = 0$, $\forall\mathbf{x}^*\in\mathbf{X}^*$ and the value of $\sigma_{pr}$ is a constant for all test points. If the prior and posterior distributions are the same, according to \eqref{KLDeq}, the KL divergence value will be zero, and the new observation does not give any additional information about the trajectory, compared to the prior assumption. However, a very small value of ${\sigma^2_{po}}^{\nabla^2}$ compared $\boldsymbol{\mu}_{po}^{\nabla^2}$ implies a high confidence of the model that the value of the Laplacian at the corresponding state is close to $\boldsymbol{\mu}_{po}^{\nabla^2}$, resulting in a large KL divergence, according to \eqref{KLDeq}. Hence, KL divergence evaluated at a state, at any given time, is proportional to its informativeness, and in our context, can be considered as a measure of the influence of a state on the trajectory.

Let $L = \{l_1, l_2, \cdot\cdot\cdot, l_Q\}$, where $l_i = \mathbb{E}[\nabla^2\phi(\mathbf{x}_i^*)] = \boldsymbol{\mu}_{i,po}^{\nabla^2}$, be the set of inferred mean of Laplacian \eqref{LaplacianDef}, at test points $\mathbf{X}^*$, and, $S_L = \{s_1, s_2, \cdot\cdot\cdot, s_Q\}$ represent the corresponding sign of Laplacian, i.e., $s_i = \frac{l_i}{|l_i|}\in \{-1,+1\}$. Let $K = \{k_1, k_2, \cdot\cdot\cdot, k_Q\}, k_i\in \mathbb{R}_+$ be the KL divergences of test points computed using \eqref{KLDeq}. Further, we define $K^a = \{k^{a}_1, k^{a}_2, \cdot\cdot\cdot, k^{a}_Q\} \in \mathbb{R}_+^Q$ and $K^r = \{k^{r}_1, k^{r}_2, \cdot\cdot\cdot, k^{r}_Q\} \in \mathbb{R}_+^Q$ as \eqref{KaKr}

\begin{equation}\label{KaKr}
  k^{a}_i =
    \begin{cases}
      k_i & \text{if $s_i = +1$}\\
      0 & \text{if $s_i = -1$},
    \end{cases}\textnormal{ and }  k^{r}_i =
    \begin{cases}
      0 & \text{if $s_i = +1$}\\
      k_i & \text{if $s_i = -1$}.
    \end{cases}          
\end{equation}

In the following, we assume a single attractor and repeller for the potential field. The extension to multiple attractors and repellers is straightforward, as long as the number of attractors and repellers are known a priori. The mean vectors and covariance matrices of the attractors and repellers are computed as weighted mean and covariances, as in \eqref{weigtedMu} and \eqref{weigtedVar}.
\begin{equation}\label{weigtedMu}
    \boldsymbol{\mu}_a = \frac{\sum_{i = 1}^Qk^a_i\mathbf{x}_i^*}{\sum_{i = 1}^Q k^a_i}
\textnormal{, }
    \boldsymbol{\mu}_r = \frac{\sum_{i = 1}^Qk^r_i\mathbf{x}_i^*}{\sum_{i = 1}^Q k^r_i}.
\end{equation}

\begin{equation}\label{weigtedVar}
    \begin{split}
        \Sigma_a = \frac{\sum_{i = 1}^Qk^a_i\left(\mathbf{x}_i^*-\boldsymbol{\mu_a}\right)\left(\mathbf{x}_i^*-\boldsymbol{\mu_a}\right)^T}{\sum_{i = 1}^Q k^a_i}
        \textnormal{, } \\
        \Sigma_r = \frac{\sum_{i = 1}^Qk^r_i\left(\mathbf{x}_i^*-\boldsymbol{\mu_r}\right)\left(\mathbf{x}_i^*-\boldsymbol{\mu_r}\right)^T}{\sum_{i = 1}^Q k^r_i}.
    \end{split}
\end{equation}
$\boldsymbol{\mu}_a$, in \eqref{weigtedMu},  is the mean attractor, which is the expected stable state towards which the price trajectory will converge, and $\Sigma_a$, in \eqref{weigtedVar}, is the corresponding covariance matrix that characterizes the uncertainty in the inferred stable state of the trajectory. Similarly, $\boldsymbol{\mu}_r$ and $\Sigma_r$, represent the expected unstable point and the associated uncertainty, respectively.

\section{Numerical Result}\label{Result} 
The organization of this section is as follows: Sec.~\ref{dataset} presents a comprehensive overview of the employed dataset. Further, in Sec.~\ref{Validation},  we validate the applicability of potential field approach for cryptocurrency market analysis using Lyapunov Stability analysis. In Sections \ref{2D_analysis} and \ref{3D_analysis}, 
we conduct a structural analysis of the cryptocurrency market using the potential field approach in Sec.~\ref{PropMeth}. We show that the structural analysis of the cryptocurrency market is particularly useful during cryptocurrency crash periods.  Section \ref{WaveletCompar} presents a comparison of the potential field approach with the popular wavelet coherence technique. Further, in Sec.~\ref{LSTMComp}, we illustrate how the inferred mean attractor can be used to improve the prediction performance of deep learning models.

\subsection{Dataset Description}\label{dataset}
The historical data of top ten cryptocurrencies in terms of market shares, namely, Bitcoin (BTC), Ethereum (ETH), XRP, Cardano (ADA), Algorand (ALGO), Binance coin (BNB), Polkadot (DOT), Litecoin (LTC), Polygon (MATIC) and Solana (SOL), has been obtained from \emph{coinmarketcap.com}, an open-source website that provides six different features of each cryptocurrency, sampled every 5 minutes. The features include opening time, opening price, closing price, highest price, lowest price, closing price, and volume of transactions within a time frame of 5 minutes, with all prices denominated in USD. 
    \begin{table}[h]\centering
        \caption{Historic BTC crashes between 2017 and 2021\label{TableCrashHist}}
        \scriptsize
        \begin{tabular}{|p{0.032\linewidth}|p{0.25\linewidth}|p{0.1\linewidth}|p{0.09\linewidth}|p{0.09\linewidth}|p{0.09\linewidth}|}
                         \hline
                         \textbf{S No.}&\textbf{Date range of Crash}&\textbf{Duration - Days}&\textbf{Max value of Bitcoin (USD)}&\textbf{Min value of Bitcoin (USD)}&\textbf{\% decline in price}\\ [0.5ex]   
                         \hline 
        1&2 to 15 Sep 2017&13 &4939.02&3187.11&35.47\\ \hline
        2&8 to 12 Nov 2017&4&7690&6100.03&20.67\\ \hline
        3&6 to 20 Jan 2018&19&17173.97&9055.03&47.27\\ \hline
        4&26 Jun to 12 Jul 2019&16&13964.64&9754.25&30.15\\ \hline
        5&13 Apr to 9 May 2021&26&64702.02&47135.68&27.15\\ \hline
        6&7 to 9 Sep 2021&2&52866.65&43088.74&18.5\\ \hline
        \end{tabular}  
    \end{table}

\begin{figure*}[ht]
\vspace{-1em}
  \centering
  \begin{minipage}[t]{0.31\textwidth}
    \includegraphics[width=\textwidth]{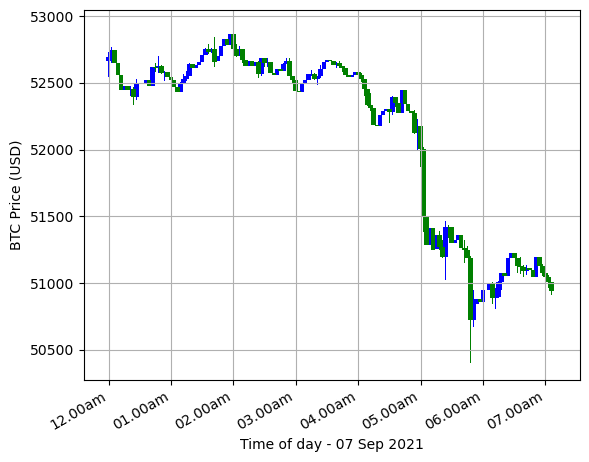}
    \caption{{Candle Stick diagram of Bitcoin from 12:00 am to 7:00 am on 7 Sep 2021. The sharp drop in BTC price from 52500 USD to 50500 USD between 4.30 am and 5.40 am is a typical instance of extreme price fluctuations of cryptocurrency}}
    \label{CandleStickBTC}
  \end{minipage}
  \hfill
  \begin{minipage}[t]{0.33\textwidth}
    \includegraphics[width=\textwidth]{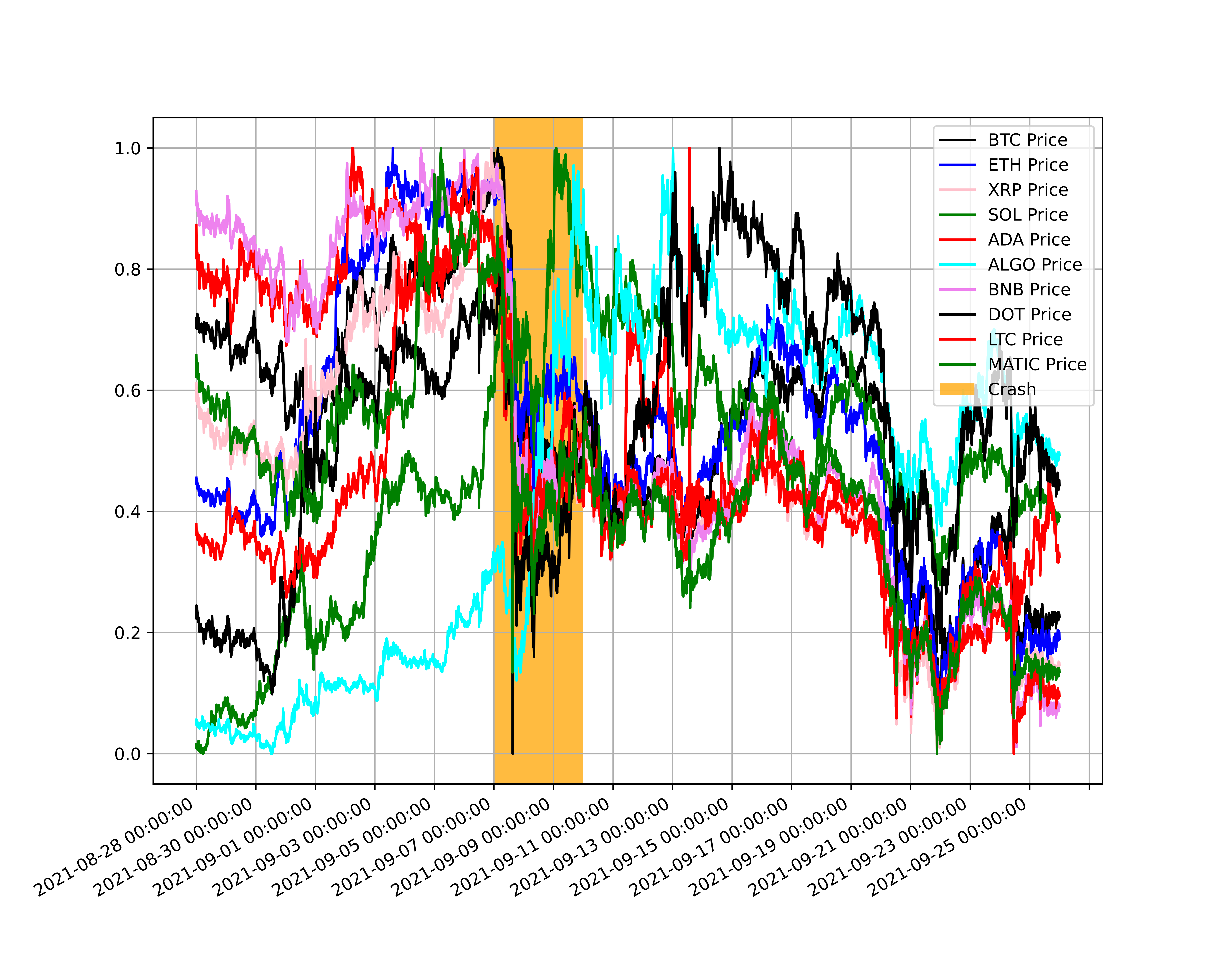}
    \caption{{  Historic prices of 10 different cryptocurrencies from 21 Aug to 25 Sep 2021. The BTC crash duration is marked in orange. Prices of cryptocurrencies are individually normalized between 0 and 1 during the observation window.}}
    \label{Actual10Cryp}
  \end{minipage}
  \hfill
  \begin{minipage}[t]{0.32\textwidth}
    \includegraphics[width=\textwidth]{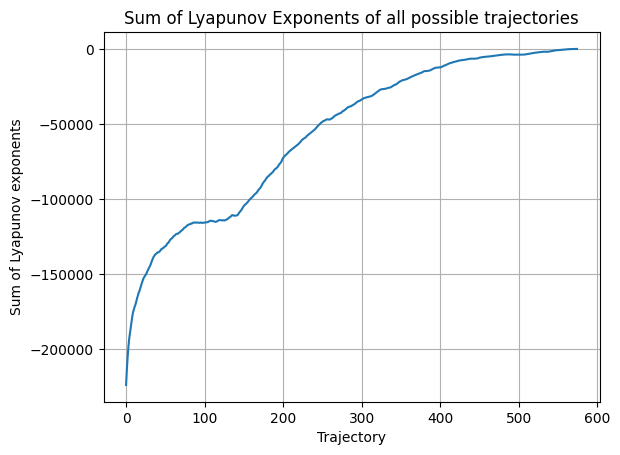}
    \caption{{  Lyapunov exponents of trajectories during 72-hour Bitcoin Crash window - April 2021. }}
    \label{Fig:Lya}
  \end{minipage}
  \hfill
\end{figure*}

In the subsequent sections, we illustrate how the potential field approach is beneficial for analysis of cryptocurrency market during “crash periods”. In this paper, we define a crash of a cryptocurrency as a rapid and significant drop in its market value, typically measured in terms of percentage change in price. A $15\%$ or more decline within a period of 24 hours or a few days is commonly defined as a crash \cite{CrashDef}. For example, the cryptocurrency market crash in late 2017, particularly affecting Bitcoin, was caused by major hacks in Korea and Japan and rumours of Bitcoin bans by certain countries, leading to a loss of confidence. The 2021 crash was primarily due to China's crackdown on Bitcoin mining and Elon Musk's announcement that Tesla would no longer accept Bitcoin, both contributing to a significant drop in Bitcoin's value. (Source: \cite{CrashReason})
During crash periods, it is well known that the volatility of the cryptocurrency market is high \cite{CrashDef} and investors tend to diversify their portfolio for profit maximization and long term financial stability \cite{correl2023WCGARCH}. To simplify analysis, we focus on crash periods of BTC, the cryptocurrency with the largest market cap, for the numerical results. The details of historic BTC crashes from the year 2017 to 2021 are listed in Table \ref{TableCrashHist}. In this paper, we have presented the results of the crash durations of $6$ to $20$ January $2017$, $13$ April to $9$ May $2021$, and $7$ to $9$ September $2021$, mentioned in the Table \ref{TableCrashHist}. Variation in historic prices of various cryptocurrencies around September 2021 BTC crash duration are shown in Fig.~\ref{Actual10Cryp}. The three-day crash window, from 7 to 9 September 2021 are shaded in orange in these figures. Further, Fig.~\ref{CandleStickBTC} shows the 7-hour candlestick plot of BTC, from 12.00am to 07.00am, on 07 September 2021. The sharp drop in BTC price from 52500 USD to 50500 USD between 4.30 am and 5.40 am is a typical instance of extreme price fluctuations of cryptocurrency. 

\subsection{Existence of a valid potential field for the cryptocurrency market}\label{Validation}
In this section, we investigate the validity of the potential field model in~\eqref{fxphirel}, using Lyapunov stability analysis (Algorithm~$1$) and Recurrence quantification Analysis (RQA)\footnote{The details of the RQA analysis are provided in supplementary material.}~\cite{RQA}.   
As the inferences regarding Lyapunov stability are identical in all cases, we present only the results for the $72$-hour observation period during the Bitcoin Crash, from $8$ to $10$~September~$2021$. 
Fig.~\ref{Fig:Lya} displays the Lyapunov exponents ($L$, the output of Algorithm~$1$), of the trajectory, arranged in ascending order. The maximum Lyapunov exponent, $\lambda_1$, is less than zero, indicating the presence of a time-varying, nonlinear potential function with an attractor~\cite{LyapunovProof,LyapExpData}. The results for the other crash durations in Table~\ref{TableCrashHist} is available in the supplementary material. 
 
In the subsequent sections, we use the inferred mean attractor and the associated uncertainty measure to analyse the market trends and structural dependencies during various BTC crash durations.

\subsection{Cryptocurrency Market Analysis - Bitcoin and Ethereum}\label{2D_analysis}
For the convenience of visualization, we start with BTC and ETH, the two major cryptocurrencies in terms of market cap.
\begin{figure*}[ht]
\vspace{-1em}
  \centering
  \begin{minipage}[t]{0.32\textwidth}
    \includegraphics[width=\textwidth]{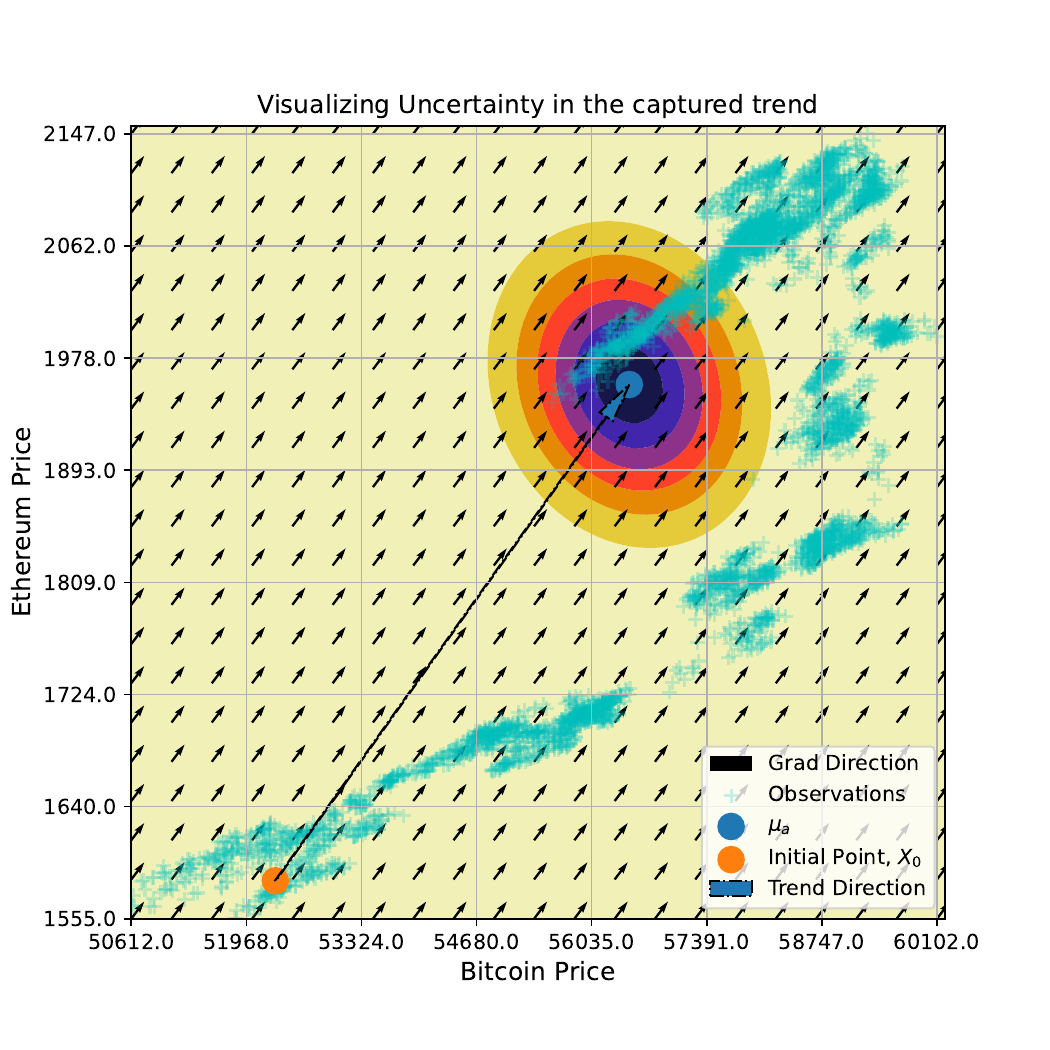}
    \caption{{Market trend captured using the proposed approach before crash period (25 March to 9 April 2021)}}
    \label{BCT2D}
  \end{minipage}
  \hfill
  \begin{minipage}[t]{0.32\textwidth}
    \includegraphics[width=\textwidth]{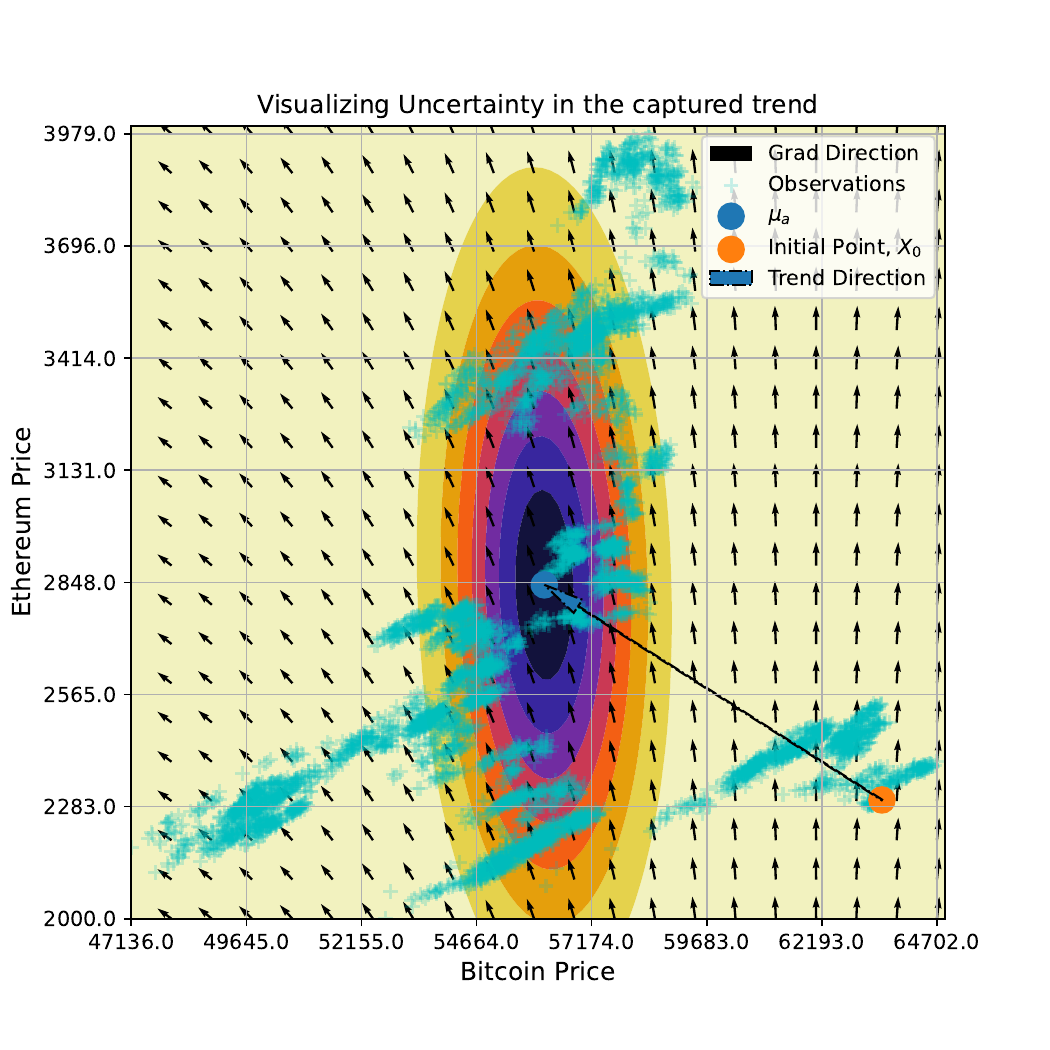}
    \caption{{Market trend captured using the proposed approach during crash period (13 April to 19 May 2021)}}
    \label{DCT2D}
  \end{minipage}
  \hfill
  \begin{minipage}[t]{0.32\textwidth}
    \includegraphics[width=\textwidth]{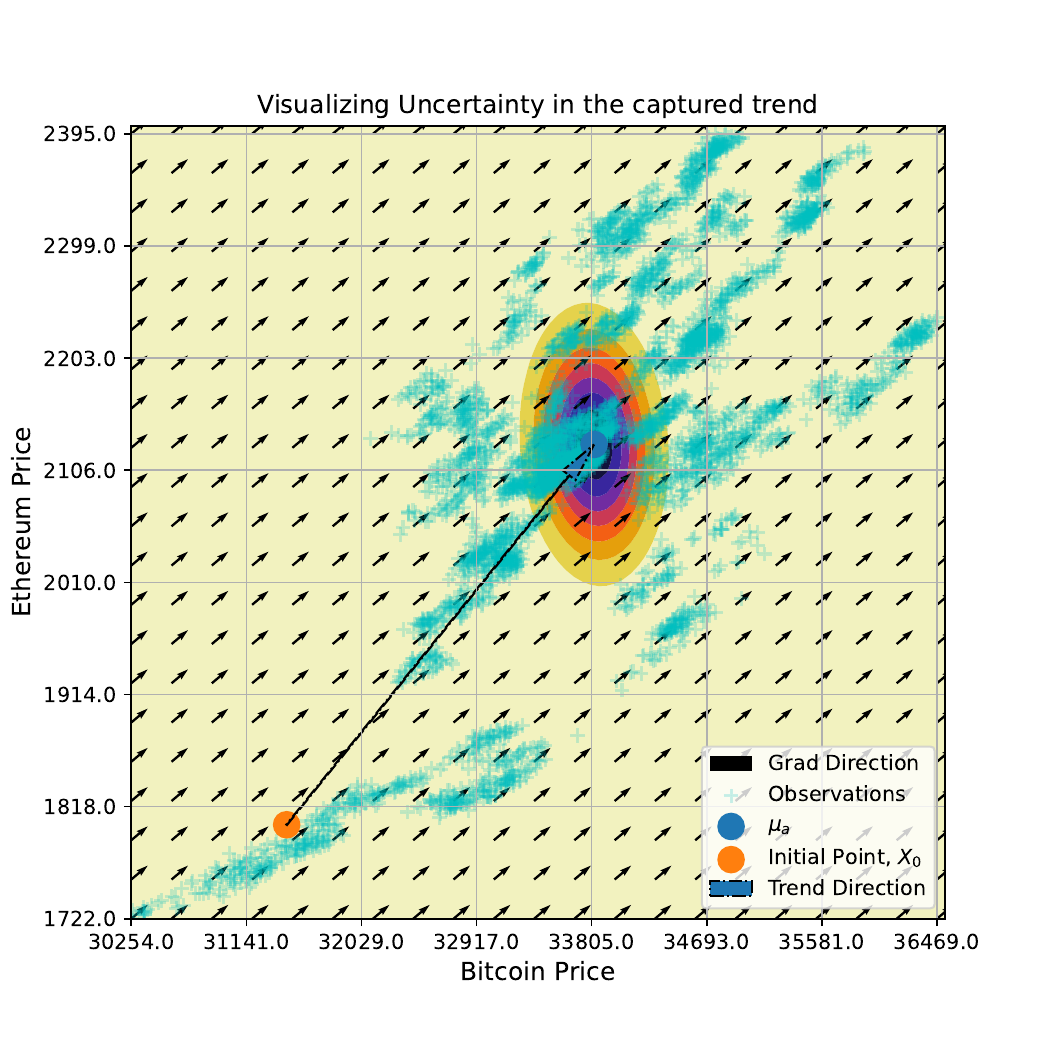}
    \caption{{Market trend captured using the proposed approach after crash period (26 June to 14 July 2021)}}
    \label{ACT2D}
  \end{minipage}
\end{figure*}
Figures \ref{BCT2D}, \ref{DCT2D}, and \ref{ACT2D}, where the horizontal axes represent the USD prices of Bitcoin, while the vertical axes represent the USD prices of Ethereum, shows the structural dependencies, obtained through the potential field approach in Sec.~\ref{PropMeth}, of BTC and ETH before (25 March 2021 to 09 April 2021), during (13 April 2021 to 09 May 2021), and after (21 June 2021 to 07 July 2021) the April 2021 Bitcoin crash. The GP in \eqref{GP} was trained\footnote{The training process of GP involves finding the values of the hyperparameters and the noise variance that best fit the training data. This is done by maximizing the log marginal likelihood of the data given the hyperparameters.} using the price trajectories of the respective windows, and the potential gradient and its Jacobian were inferred in a finite mesh grid of test points in the state space using \eqref{GPosMu} and \eqref{JacMu}, respectively, for all windows. 
The vector lines corresponding to the inferred posterior mean of potential gradient, as in \eqref{GPosMu}, in the mesh grid, are shown as arrows in the state space in Figures \ref{BCT2D}, \ref{DCT2D}, and \ref{ACT2D}. 
Further, \eqref{LaplacianDef} was used to compute the Laplacian at the test points. 
Subsequently, the KL divergence between the prior and posterior distributions of the Laplacian at all test points are evaluated using \eqref{KLDeq}. Finally, the mean of the mean attractor ($\boldsymbol{\mu}_a = [\mu_{BTC},\mu_{ETH}]^T$) and covariance matrix ($\Sigma_a$) in the respective windows, were obtained using~\eqref{weigtedMu} and~\eqref{weigtedVar}. Cyan markers on the state space indicate the states of the observed price trajectory, during the respective observation windows. 
The initial state of an observation window will be denoted as $\mathbf{x}_0 = [x_{BTC,0}, x_{ETH,0}]^T$, and is marked in orange. The blue markers represent the posterior mean of the inferred mean attractors of each window. The concentric ellipses, the contours\footnote{Contours of a multivariate Gaussian probability density function, are lines or surfaces that connect the points of equal probability density. The contours of this distribution will be ellipses if the covariance matrix is positive semidefinite} of the Gaussian cloud $\mathcal{N}(\boldsymbol{\mu}_a,\Sigma_a)$, represent the uncertainty in the inference of mean attractor.

The structural dependency inferences related to cryptocurrency market such as market trend, volatility and its associated uncertainty, convergence and market stabilization, and visualizing shift in market trends are discussed in Sec.~\ref{TrendLines} to Sec.~\ref{temporal}.
\medskip
\subsubsection{The Trend Lines}\label{TrendLines}
Trend Line, a directed line segment from the initial state to the mean attractor, in Figures \ref{BCT2D}, \ref{DCT2D}, and \ref{ACT2D}, provides insights into the trends in price corrections, the overall correlation during the observation window and the possible direction of asset shift, between BTC and ETH. The magnitude and inclination of trend lines, in all the crash durations of interest, are quantified in Table~\ref{Fig:Table_Windows}. 

Before the crash of April 2021, the Bitcoin price showed a rapid increase, which resulted in a bubble\footnote{When an asset is traded at a price exceeding that asset's intrinsic value, mostly due to speculations. Speculation in trading terminology refers to the investment in the market, in the hope of gain but with the risk of loss. Further, independent of its market price, the intrinsic value of a cryptocurrency is determined by its utility, scarcity and technological application.\cite{CrashDef}} formation, as characterized by a trend line inclination of $\sim60$ degrees\footnote{percentage positive price correction of BTC is greater than that of ETH}, in Fig.~\ref{BCT2D}. During the crash, the bubble burst, leading to a substantial negative correction in the Bitcoin price, which fell on an average  by $-10000 USD$, characterized by $\sim135$ degrees inclination of trend line in Fig.~\ref{DCT2D}. The Bitcoin price started showing positive corrections during the recovery period after the crash, as depicted in Fig.~\ref{ACT2D}. 
 
In addition, it can be observed from Table~\ref{Fig:Table_Windows} that the price of Ethereum (ETH) has undergone positive price corrections of approximately $+350USD$, $+650USD$, and $+300USD$, respectively, before, during and after the BTC crash of April 2021. Notably, during the Bitcoin (BTC) market downturn, there was a substantial high positive correction observed in the price of Ethereum, as shown in Fig.~\ref{DCT2D}. This trend suggests a possible asset shift from Bitcoin to Ethereum during the Bitcoin crash period, whereby individuals exchanged their BTC for ETH. This shift in asset allocation increased the demand for ETH, leading to a favourable effect on its price. 
\medskip
\subsubsection{Uncertainty in mean attractor and confidence of price correction} 
The \emph{uncertainty} in the inferred mean attractor is characterized by the Gaussian cloud, $\mathcal{N}(\boldsymbol{\mu}_a, \Sigma_a)$, contours of which in state space are represented as concentric ellipses centred at the mean attractors, in Figures \ref{BCT2D}, \ref{DCT2D}, and \ref{ACT2D}. Here, $\boldsymbol{\mu}_a$ and $\Sigma_a$ are obtained using \eqref{weigtedMu} and \eqref{weigtedVar}, respectively. The multivariate Gaussian probability density function $\mathcal{N}(\boldsymbol{\mu}_a, \Sigma_a)$ quantifies the likelihood of the inferred mean attractors being located within this Gaussian cloud.

We can infer the confidence of price correction (whether its an uptrend or downtrend of price) as the probabilities of positive and negative price corrections of the cryptocurrencies, starting from the initial state, $\mathbf{x}_0$, based on the inferred distribution $\mathcal{N}(\boldsymbol{\mu}_a, \Sigma_a)$.
The probability of positive price correction for BTC is computed as $P(X_{BTC} > x_{BTC,0})$, as in \eqref{ProbCorrec}, 
\begin{equation}\label{ProbCorrec}
    \begin{split}
        P(X_{BTC} > x_{BTC,0})& 
        = \int_{x_{BTC,0}}^{\infty} f_{X_{BTC}}(x)dx,
    \end{split}     
\end{equation}
where, $f_{X_{BTC}}(x)$ is the probability density function of BTC price and $X_{BTC}\sim\mathcal{N}(\mu_{BTC}, \sigma_{BTC})$. The probability of negative correction is computed as~{$1 - P(X_{BTC} > x_{BTC,1})$}. Similarly, the probabilities  pertaining to the ETH price corrections are also computed. These findings are quantified and summarized in Table~{\ref{Fig:Table_Windows}}. Throughout all crash periods, the price of BTC exhibited notable positive corrections prior to the crash, which led to the formation of a speculative bubble. Subsequently, a sudden negative correction ensued during the crash itself, ultimately followed by a gradual positive correction during the recovery phase. It is evident from the results in Table~{\ref{Fig:Table_Windows}} that the inferred mean attractor correctly characterizes the price correction trends, before, during and after various crash durations, with a high degree of confidence. 

    \begin{table*}[htbp]
      \centering
      \caption{Characterizing trends and uncertainties – BTC and ETH}
      \begin{tabular}{|p{1.3cm}|p{0.9cm}|p{1.5cm}|p{1.3cm}|p{1.3cm}|p{0.7cm}|p{0.7cm}|p{0.7cm}|p{0.7cm}|}
        \hline
        \multicolumn{1}{|p{1.3cm}|}{\textbf{Crash Window}} & \multicolumn{1}{p{0.9cm}|}{} &\multicolumn{2}{p{2.8cm}|}{\textbf{Trend Line Magnitude (USD)}} & \multicolumn{1}{p{1.4cm}|}{\textbf{Trend Line Direction (Degrees)}} & \multicolumn{2}{p{1.4cm}|}{\textbf{Probability of Negative Correction}} & \multicolumn{2}{p{1.4cm}|}{\textbf{Probability of Positive Correction}}  \\
        \cline{3-4} \cline{6-7} \cline{8-9}
        \multicolumn{1}{|p{1.3cm}|}{} & \multicolumn{1}{p{0.9cm}|}{} & \multicolumn{1}{p{1.5cm}|}{\textbf{BTC}} & \multicolumn{1}{p{1.3cm}|}{\textbf{ETH}} & \multicolumn{1}{p{1.4cm}|}{} & \multicolumn{1}{p{0.7cm}|}{\textbf{BTC}} & \multicolumn{1}{p{0.7cm}|}{\textbf{\textbf{ETH}}} & \multicolumn{1}{p{0.7cm}|}{\textbf{BTC}} & \multicolumn{1}{p{0.7cm}|}{\textbf{ETH}} \\
        \hline
        &Before &2893.32 &125.429 &40 &0 &0 &1 &1 \\
        \cline{2-9}
        Dec 2017 &During &-3468.254 &281.435 &135 &1 &0 &0 &1 \\
        \cline{2-9}
        &After &2337.132 &223.258 &50 &0 &0 &1 &1 \\
        \hline
        &Before &4168.778 &373.964 &60 &0 &0 &1 &1 \\
        \cline{2-9}
        \textbf{Apr 2021} &During &-7359.387 &542.325 &135 &1 &0 &0 &1 \\
        \cline{2-9}
        &After &2371.57 &326.296 &45 &0 &0 &1 &1 \\
        \hline
        &Before &293.305 &21.645 &45 &0 &0 &1 &1 \\
        \cline{2-9}
        Sep 2021 &During &-229.072 &76.894 &120 &1 &0 &0 &1 \\
        \cline{2-9}
        &After &-79.729 &-7.862 &218 &0 &0 &1 &1 \\
        \hline
      \end{tabular}  
      \label{Fig:Table_Windows}
    \end{table*}
\medskip
\subsubsection{Characterizing the volatility and correlation of cryptocurrencies}\label{StrDepPot} 

The contours of the posterior covariance of the mean attractor is a hyper-ellipsoid, the principal axes of which can give insights about the volatility of cryptocurrencies and the correlation between individual components of the price vector~\cite{murphy2012machine}. In 2-dimensional case, as in Figures~\ref{BCT2D} and \ref{ACT2D}, involving BTC and ETH, when the principal axis and the horizontal axis (BTC price) form an acute angle\footnote{In Figures~\ref{BCT2D}, \ref{DCT2D}, and \ref{ACT2D}, different scales are used for horizontal and vertical axes. Refer Table~\ref{2DStructDep}, for exact inclination of principal axis.}, it indicates a positive correlation between Bitcoin (BTC) and Ethereum (ETH). Conversely, as in Fig.~\ref{DCT2D}, an obtuse angle suggests a negative correlation between the cryptocurrencies. Horizontally, or vertically oriented principal axis, suggest no correlation between the individual components of the price vector. This orientation can be obtained from the principal component (the eigen vector corresponding to the largest eigen value) of the covariance matrix $\Sigma_a$ \cite{murphy2012machine}. Furthermore, the length of principal axis, which is the largest eigen value of $\Sigma_a$, being proportional to the dispersion in price, can quantify the volatility of cryptocurrencies. The principal components and the  corresponding eigen values for various crash durations are presented in Table~\ref{2DStructDep}.

Several aspects related to the volatility behaviour of BTC and ETH can be inferred from Figures~\ref{BCT2D}, \ref{DCT2D}, and \ref{ACT2D}, and Table~\ref{2DStructDep}. 
The volatility increases during crash periods, as quantified by the eigen values corresponding to the principal component of the covariance matrix ($\Sigma_a$). From Table~\ref{2DStructDep}, we can note that the eigenvalues exhibit a change in value, specifically $0.047$ and $0.069$, correspondingly, prior to ($25$ March to $9$ April, $2021$) and during ($13$ April to $9$ May, $2021$) the April $2021$ BTC crash. This change indicates a significant increase in volatility during the April $2021$ crash period.  Further, after the April $2021$ crash duration ($26$ June to $14$ July $2021$), the eigen value reduced to $0.067$, indicating a reduction in volatility, during the market recovery, after the April $2021$ crash. Similar results can be inferred for September $2021$ BTC crash, from Table~\ref{2DStructDep}. In contrast to the April $2021$ and September $2021$ BTC crashes, we observe that the eigenvalue increased from $0.051$ during the crash to $0.061$ after the crash of January $2018$. This increase in volatility, lies in the sustained period during which the BTC price remained at the crashed value, lasting nearly one year following the January $2018$ crash. Moreover, the recovery of the cryptocurrency market from this crash occurred only in early $2019$.

Further, it can be observed from Table~\ref{2DStructDep} that, during the period of $25$ March to $9$ April $2021$, prior to the BTC crash, the principal axis made an orientation of $17\degree$ with respect to the BTC axis. This orientation indicates a slight positive correlation between BTC and ETH, as both prices demonstrated a positive trend. In contrast, during the crash period from $13$ April to $9$ May, the principal axis and BTC axis formed an angle of $95.7\degree$, indicating a slight negative correlation between BTC and ETH. This finding aligns with the fact that while BTC experienced a significant negative correction during the crash, ETH exhibited a positive correction. Additionally, from $26$ June to $14$ July, after the crash, the principal axis exhibited an angle of $46\degree$ with the BTC axis, indicating a very high positive correlation between BTC and ETH. These observations suggest that the structural dependency between BTC and ETH increased considerably following the April $2021$ crash. Moreover, by combining the insights regarding price correction trends (Table~\ref{Fig:Table_Windows}), direction of inferred potential gradient (Figures \ref{BCT2D} through \ref{ACT2D}) and the information on structural dependence (Table \ref{2DStructDep}), it is possible to infer a potential shift of assets from BTC to ETH during the crash period of April 2021.
    \begin{table*}[!htp]\centering
        \caption{Structural dependence during convergence – BTC and ETH}\label{tab: }
        \scriptsize
        \begin{tabular}{|p{0.13\linewidth}|p{0.07\linewidth}|p{0.1\linewidth}|p{0.09\linewidth}|}
         \hline
         \textbf{Observation window} & \textbf{Eigen Value} & \textbf{Eigen Vector} & \textbf{Phase (BTC (x) and ETH (y))} \\ [0.5ex]   
         \hline 
            7 to 18 Dec 2017 &0.025 &[0.605, -0.796] &128\degree \\  [0.1ex]
            \hline
            6 to 20 Jan 2018 &0.051 &[-0.553, -0.833] &56.42\degree \\  [0.1ex]
            \hline
            6 to 20 Feb 2018 &0.066 &[0.720, 0.693] &43.88\degree \\   [0.1ex]
            \hline
            \textbf{25 Mar to 9 Apr 2021} &\textbf{0.047} &\textbf{[0.956, 0.293]} &\textbf{17.01\degree} \\  [0.1ex]
            \hline
            \textbf{13 Apr to 9 May 2021} &\textbf{0.069} &\textbf{[0.099, -0.995]} &\textbf{95.7\degree} \\   [0.1ex]
            \hline
            \textbf{26 Jun to 14 Jul 2021} &\textbf{0.067} &\textbf{[-0.691, -0.723]} &\textbf{46.33\degree} \\   [0.1ex]
            \hline
            29 to 30 Aug 2021 &0.025 &[-0.645, -0.764] &49.8\degree \\    [0.1ex]
            \hline
            7 to 9 Sept 2021 &0.031 &[0.715, 0.698] &44.3\degree \\   [0.1ex]
            \hline
            25 to 26 Sept 2021 &0.003 &[-0.660, -0.751] &48.74\degree \\  [0.1ex]
            \hline
        \end{tabular}
        \label{2DStructDep}
    \end{table*}
\medskip
\subsubsection{Convergence Characteristics and Market Stabilization}\label{ConvChar}
\begin{figure*}[htbp]
  \centering
  \begin{minipage}[t]{0.32\textwidth}
    \includegraphics[width=\textwidth]{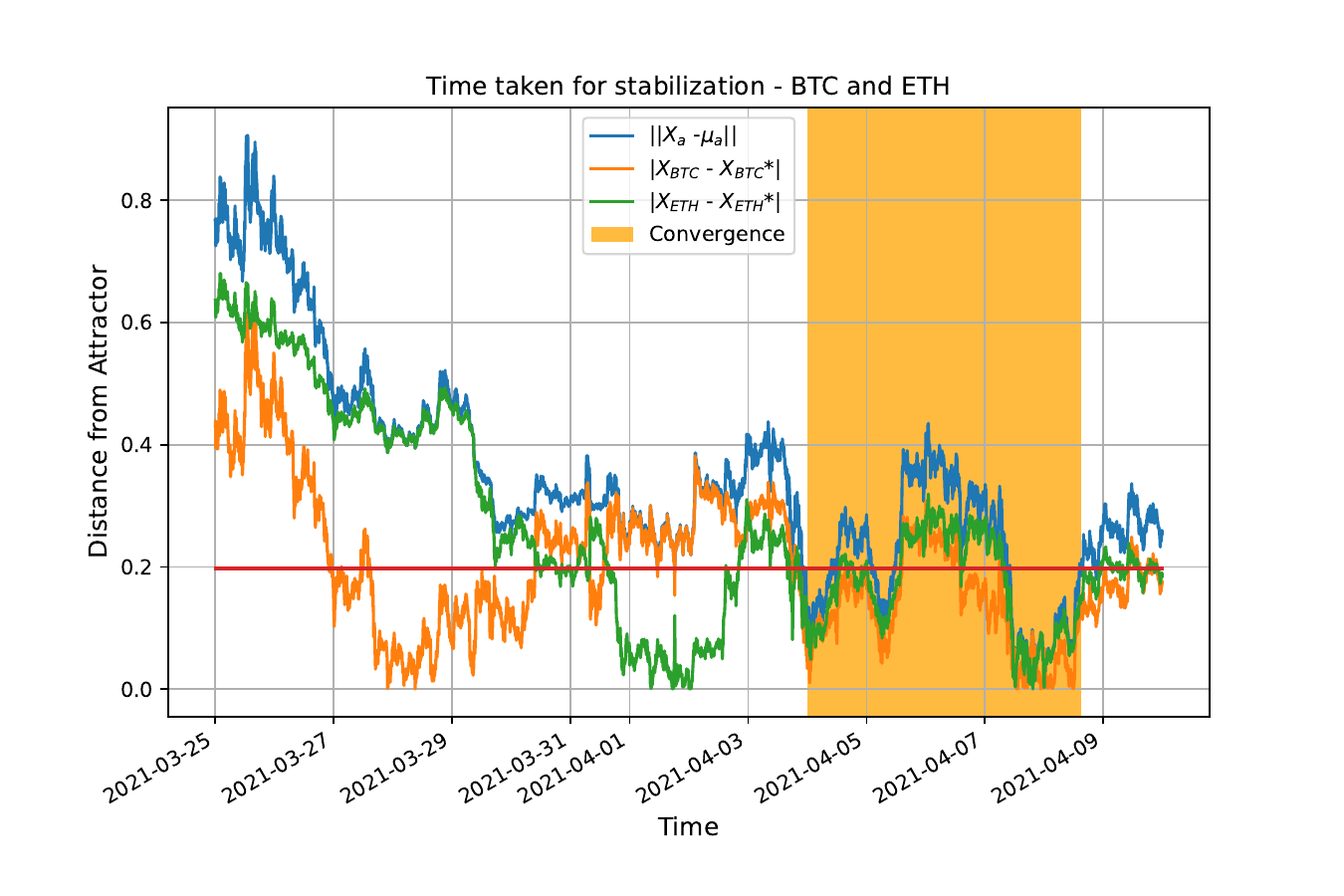}
    \caption{{Convergence to attractor - Before crash (25 March to 9 April 2021)}}
    \label{Fig:Conv2dBC}
  \end{minipage}
  \hfill
  \begin{minipage}[t]{0.32\textwidth}
    \includegraphics[width=\textwidth]{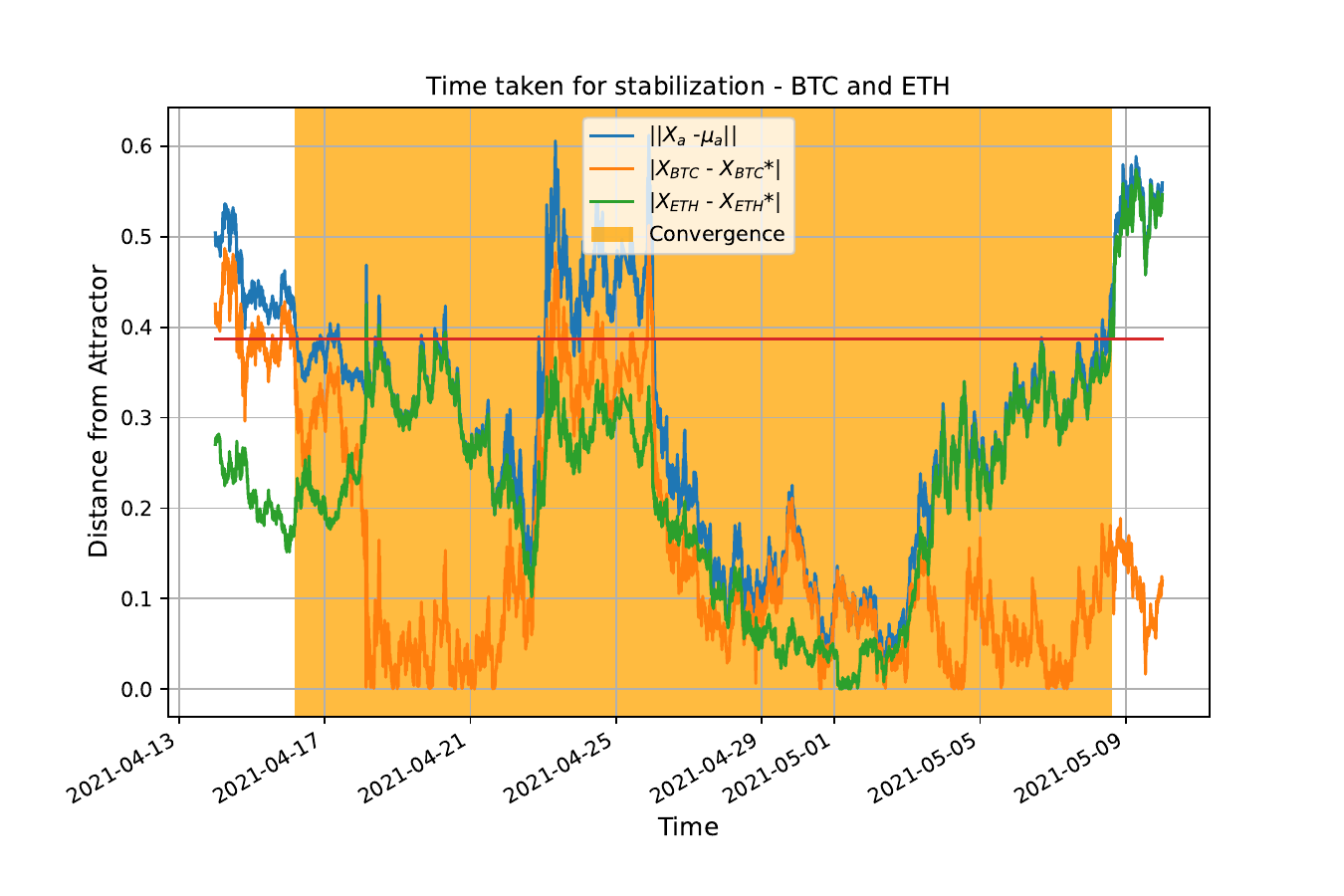}
    \caption{{Convergence to attractor - During crash (13 April to 19 May 2021)}}
    \label{Fig:Conv2dDC}
  \end{minipage}
  \hfill
  \begin{minipage}[t]{0.32\textwidth}
    \includegraphics[width=\textwidth]{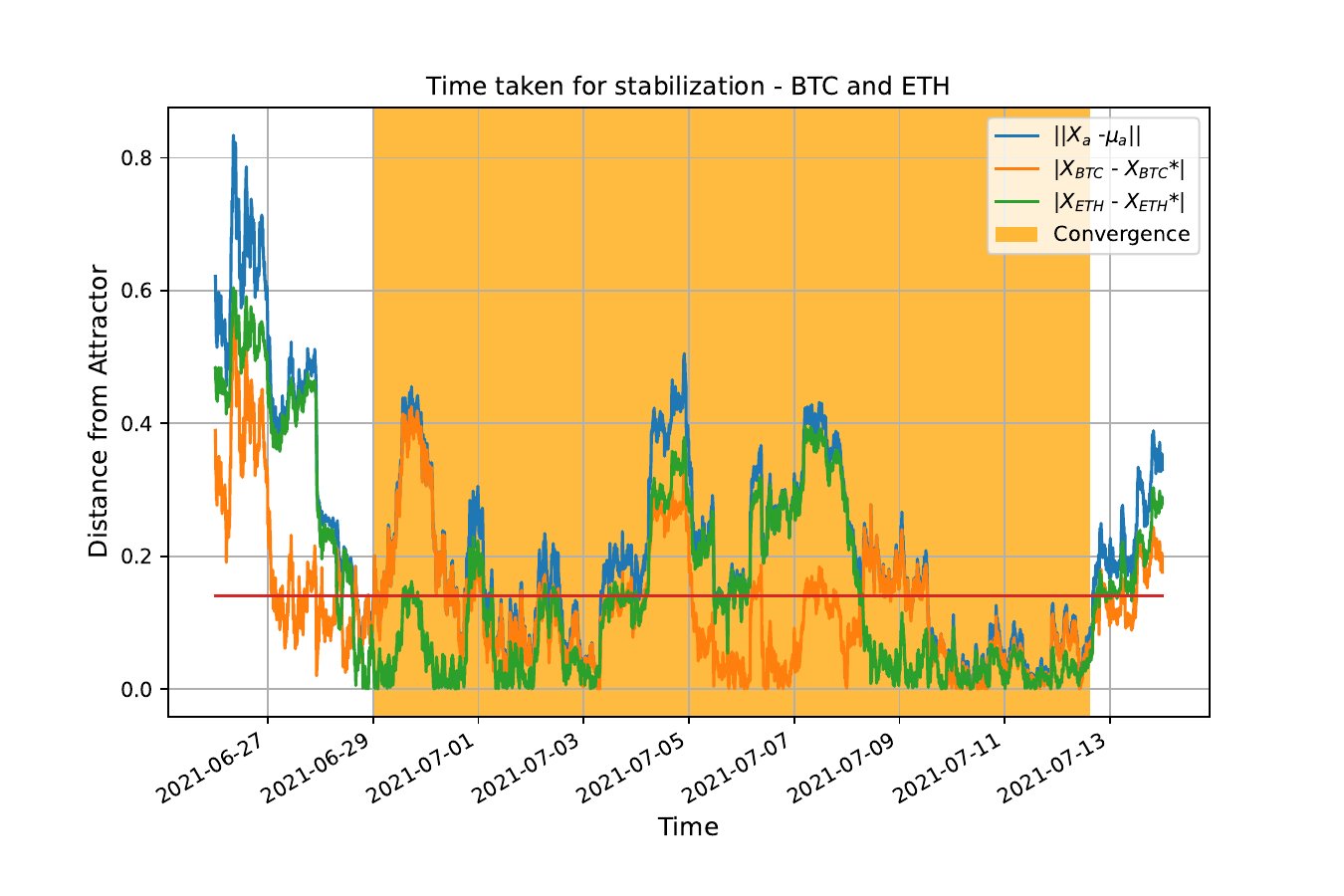}
    \caption{{Convergence to attractor – After crash (26 June to 14 July 2021)}}
    \label{Fig:Conv2dAC}
  \end{minipage}
\end{figure*}

In this section, we empirically analyse the convergence property of the cryptocurrency price trajectory towards the inferred mean attractor, and the temporal evolution of the mean attractor. The cryptocurrency market is deemed to be stabilized when the price trajectory falls within a distance of $\sigma_a$ from the mean attractor, as implied by \eqref{convergCond}. 
\begin{equation}\label{convergCond}
    ||\mathbf{x}_n - \boldsymbol{\mu}_a|| \le \sigma_a, \text{ where } n = 1,\cdot\cdot\cdot,N,
\end{equation}where, $\mathbf{x}_n$ is the state of the trajectory at time $t_n$. The quantity, $\sigma_a = |\Sigma_a|^{\frac{1}{2}}$ ($\Sigma_a$ is inferred from \eqref{weigtedVar}), represents the area of the Gaussian cloud, centred at the mean attractor, that encloses $68.27\%$ of the total probability mass of the multivariate Gaussian distribution, $\mathcal{N}(\boldsymbol{\mu},\Sigma_a)$. Figures, \ref{Fig:Conv2dBC} to \ref{Fig:Conv2dAC}, depict the convergence behaviour of cryptocurrency market dynamics over time, indicating the timeline of market stabilization. The horizontal axis represents time, while the vertical axis represents the normalized Euclidean distance from the attractor. The horizontal lines coloured in red, in Fig. \ref{Fig:Conv2dBC} to Fig. \ref{Fig:Conv2dAC}, depict the inferred uncertainty, which is the normalized standard deviation, $\sigma_a$, of the location of the mean attractor. The blue graph represents the distance between the states of trajectory from mean attractor as a function of time. The orange and green graphs represent the absolute distances of BTC price and ETH price respectively from the corresponding components of mean attractor. As a result of the market's intrinsic volatility, the price may display temporary divergences from the attractor.
However, on average, during the convergence period, the region which is highlighted in orange, in Figures \ref{Fig:Conv2dBC} through \ref{Fig:Conv2dAC}, the price is observed to return and remain within a distance of $\sigma_a$ from the mean attractor. If the Euclidean distance between the trajectory's states and the inferred mean attractor consistently exceeds $\sigma_a$, it indicates a displacement of the attractor's position caused by a shift in market trend. 

Table \ref{Fig:Convergence 2 Crypto} presents the BTC and ETH components of the inferred mean attractor, the onset and ending time of convergence and duration of convergence for different observation windows. For instance, it can be observed that during the BTC market downturn in April 2021, the price trajectory approached the mean attractor within a distance of $\sigma_a$ on April 16, 2021. The trajectory remained in this vicinity until May 5, 2021, indicating convergence towards the mean attractor. However, the trajectory deviated away from the mean attractor (inferred during the crash window), which can be attributed to the rebound in the BTC price during the first week of May 2021.

    \begin{table*}[!htp]\centering
        \caption{Convergence characteristics – BTC and ETH}\label{tab: }
        \scriptsize
        \begin{tabular}{|p{0.13\linewidth}|p{0.11\linewidth}|p{0.11\linewidth}|p{0.13\linewidth}|p{0.12\linewidth}|p{0.07\linewidth}|}
         \hline
         \textbf{Observation window} & \textbf{BTC component of mean attractor} & \textbf{ETH component of mean attractor} & \textbf{Onset of Convergence} & \textbf{End of Convergence} & \textbf{Convergence Duration} \\ [0.5ex]
         \hline
            7 to 18 Dec 2017 &16663.32 &534.42 & Dec 11 2017 & Dec 12 2017 &1 day \\
            \hline
            6 to 20 Jan 2018 &13450.745 &1236.215 & Jan 08 2018 & Jan 16 2018 &8 days \\
            \hline
            6 to 20 Feb 2018 &8898.132 &877.248 & Feb 08 2018 & Feb 14 2018 &6 days \\
            \hline
            25 Mar to 9 April 2021 &58138.71 &1925.75 & Apr 04 2021 & Apr 08 2021 &4 days \\
            \hline
            13 April to 9 May 2021 &56142 &2841 & Apr 16 2021 & May 08 2021 &22 days \\
            \hline
            26 June to 14 July 2021 &33823 &2128 & Jun 28 2021 & Jul 12 2021 &14 days \\
            \hline
            29 to 30 Aug 2021 &48554.235 &3205.435 & Aug 28 2021 & Aug 29 2021 &1 day \\
            \hline
            7 to 9 Sept 2021 &46434 &3476 & Sep 07 2021 & Sep 09 2021 &2 days \\
            \hline
            25 to 26 Sept 2021 &42589.74 &2913.388 & Sep 25 2021 & Sep 26 2021 &1 day \\
            \hline
        \end{tabular}
        \label{Fig:Convergence 2 Crypto}
    \end{table*}
\medskip
\subsubsection{Temporal advancement of attractors – Visualizing the shift in market trend}\label{temporal}
\begin{figure*}[htbp]
  \centering
  \begin{minipage}[t]{0.32\textwidth}
    \includegraphics[width=\textwidth]{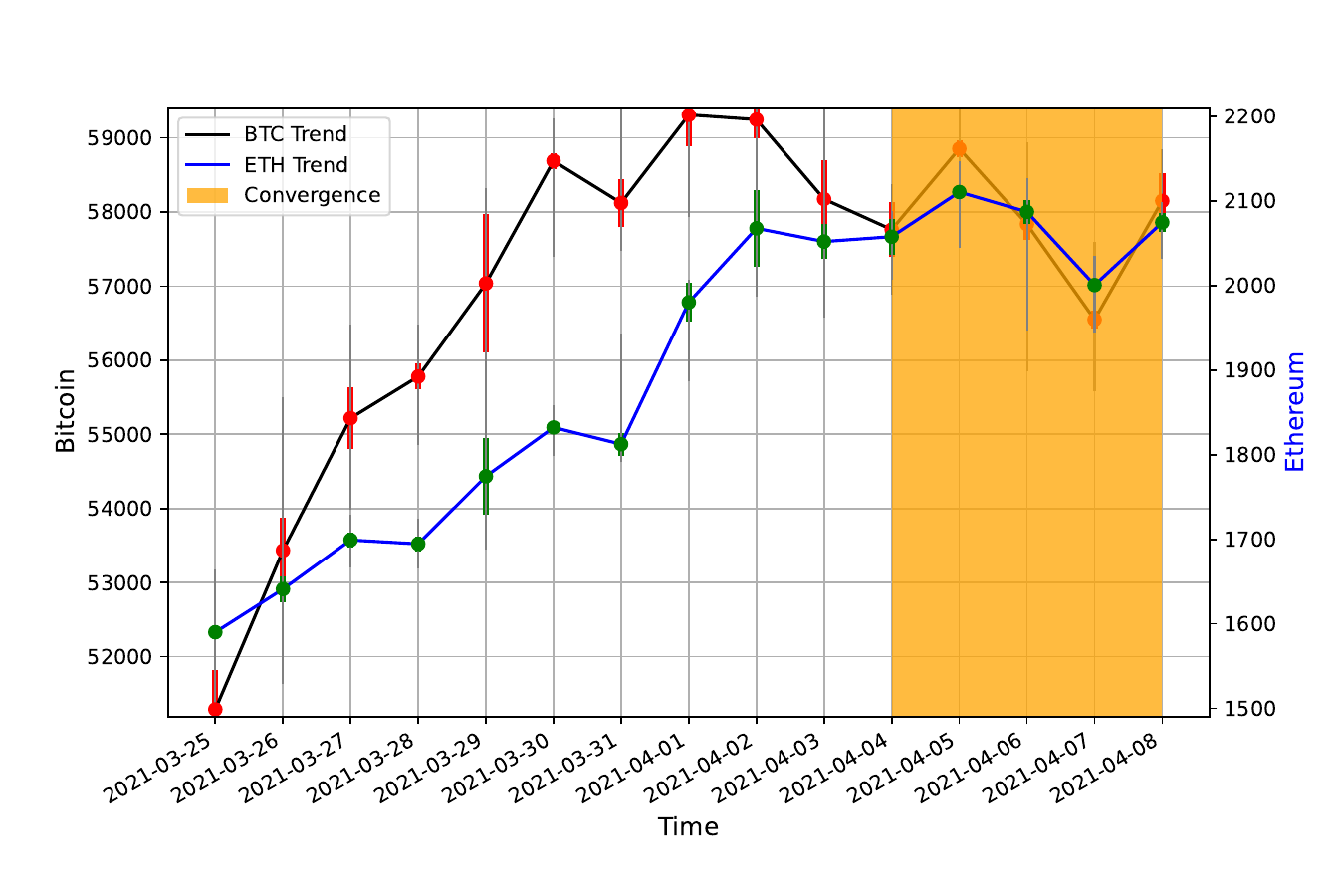}
    \caption{{  Temporal evolution of attractor – Before crash (25 March to 9 April 2021)}}
    \label{Fig:TrendChangeBC}
  \end{minipage}
  \hfill
  \begin{minipage}[t]{0.32\textwidth}
    \includegraphics[width=\textwidth]{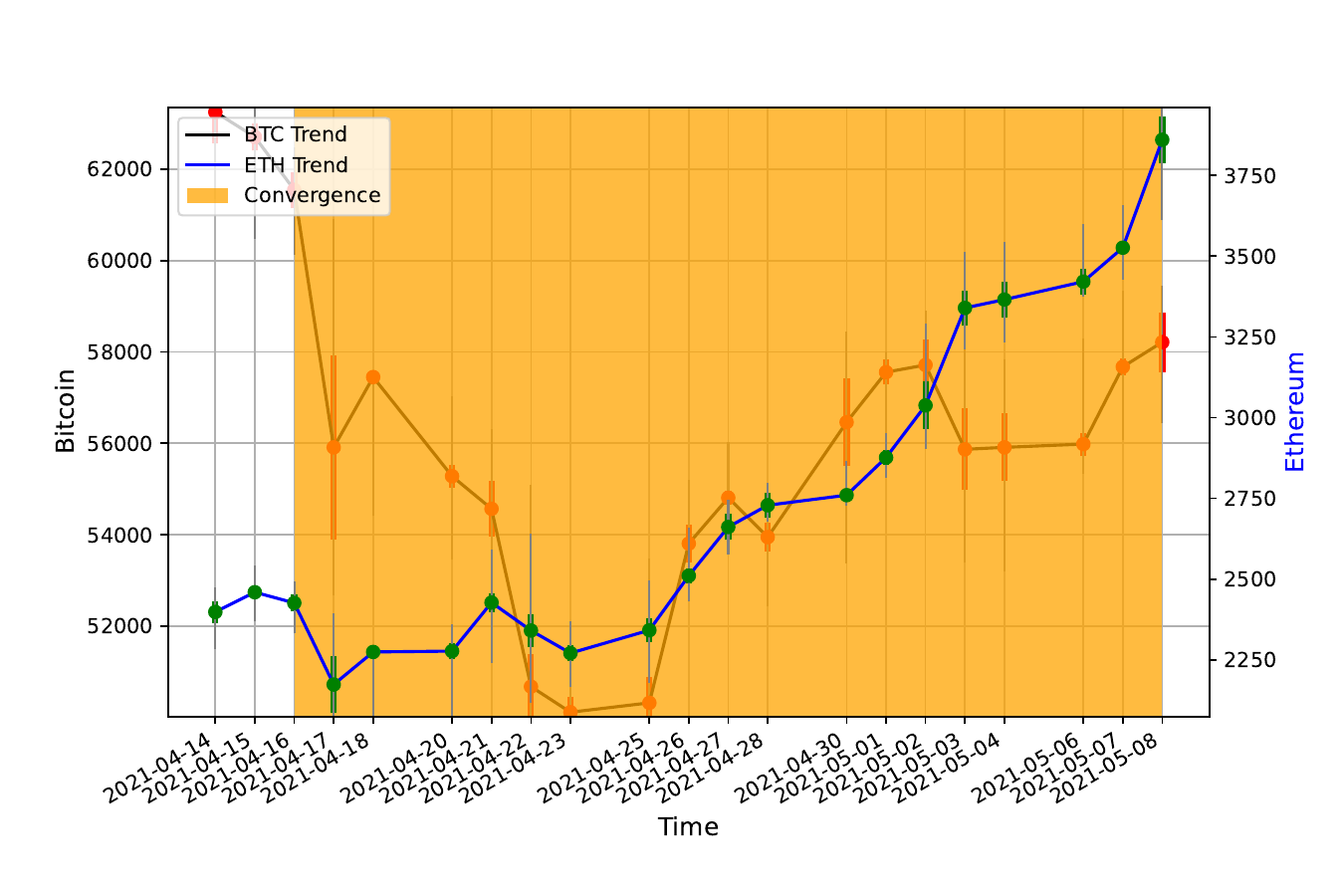}
    \caption{{  Temporal evolution of attractor – During crash (13 April to 19 May 2021)}}
    \label{Fig:TrendChangeDC}
  \end{minipage}
  \hfill
  \begin{minipage}[t]{0.32\textwidth}
    \includegraphics[width=\textwidth]{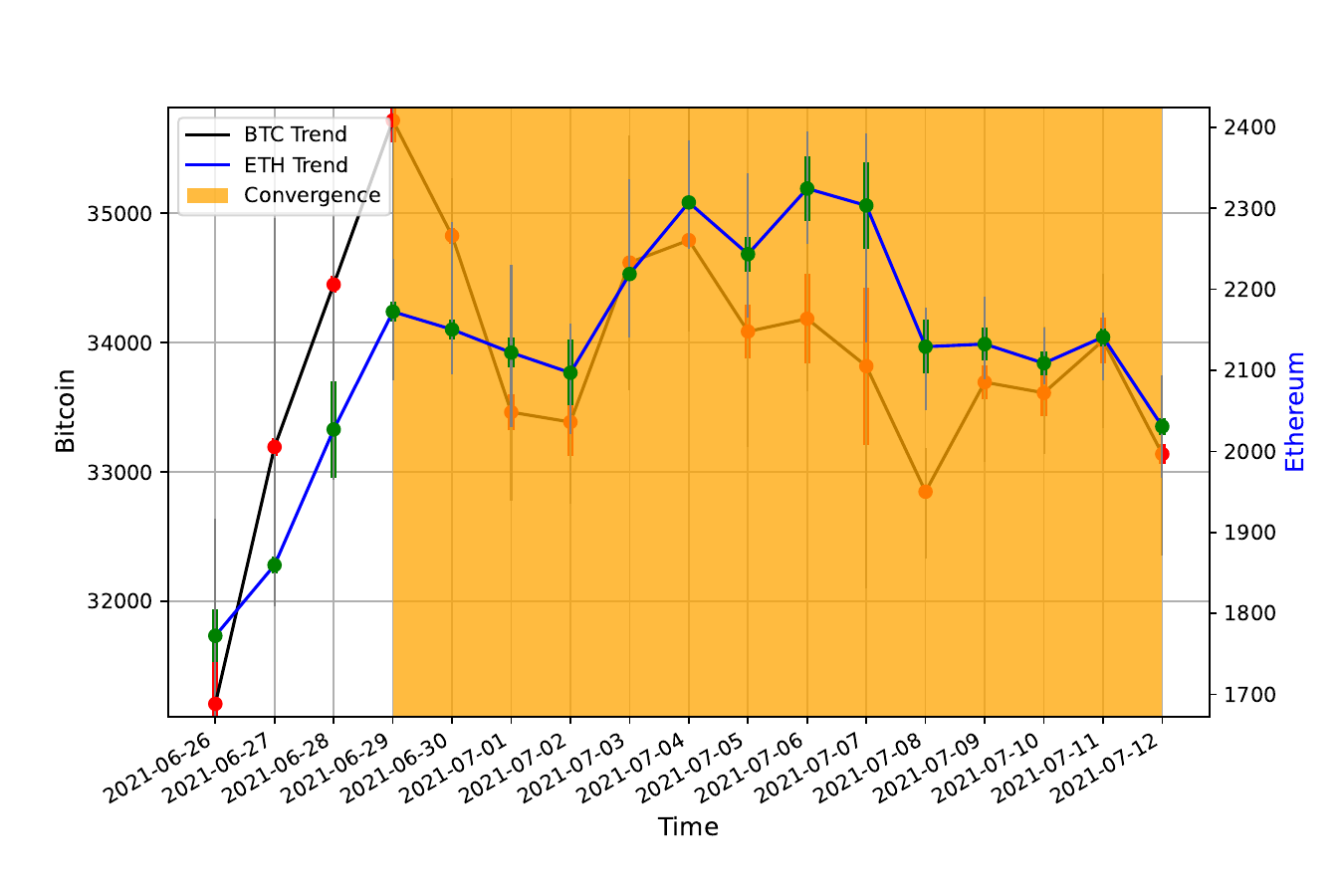}
    \caption{{Temporal evolution of attractor - After crash (26 June to 14 July 2021)}}
    \label{Fig:TrendChangeAC}
  \end{minipage}
\end{figure*}

Figures \ref{Fig:TrendChangeBC} through \ref{Fig:TrendChangeAC} depict the temporal evolution of attractors across distinct 1-day sub-windows within the observation windows presented in Figures \ref{BCT2D} through \ref{ACT2D}, respectively. The BTC and ETH components of the mean attractor, as defined by equation \eqref{weigtedMu}, are displayed separately with accompanying error plots, representing the associated uncertainties. The uncertainty, $\sigma_a = |\Sigma_a|^{\frac{1}{2}}$,  obtained using equation \eqref{weigtedVar}, over the respective 1 day subwindows. The shaded areas correspond to the region of convergence, similar to Figures \ref{Fig:Conv2dBC} through \ref{Fig:Conv2dAC}. The movement of attractors effectively captures the BTC price bubble formation prior to the crash in Fig.~\ref{Fig:TrendChangeBC}, the severe negative correction of BTC price during the crash in Fig.~\ref{Fig:TrendChangeDC}, and the recovery of BTC price after the crash in Fig.~\ref{Fig:TrendChangeAC}. Additionally, the positive price correction of ETH price during all three windows is evident from Figures \ref{Fig:TrendChangeBC} through \ref{Fig:TrendChangeAC}.

Each time interval in the cryptocurrency's observation sub-windows is represented by an error bar, which provides useful information on the corresponding volatility. The error bar captures the regions around $\mu_{BTC}$ and $\mu_{ETH}$  bounded by $\pm\sigma_{BTC}$ and $\pm\sigma_{ETH}$, respectively. An increase in the size of the error bar indicates higher uncertainty and, hence, higher volatility, while a decrease in the size of the error bar indicates lower volatility. For instance, the error bars at the mean attractor points in Figures \ref{Fig:TrendChangeBC} and \ref{Fig:TrendChangeAC} are relatively small compared to the mean value across the subwindows. However, during the window spanning from April 16 to 17, 2021, as shown in Fig.~\ref{Fig:TrendChangeDC}, a significantly wide error bar can be observed for BTC, indicating a drastic negative correction of BTC price by $16\%$ in just two days, at the onset of the BTC market downturn.

\subsection{Cryptocurrency Market Analysis – More than 2 cryptocurrencies}\label{3D_analysis}
\begin{figure*}[htbp]

  \centering
  \begin{minipage}[t]{0.32\textwidth}
    \includegraphics[width=\textwidth]{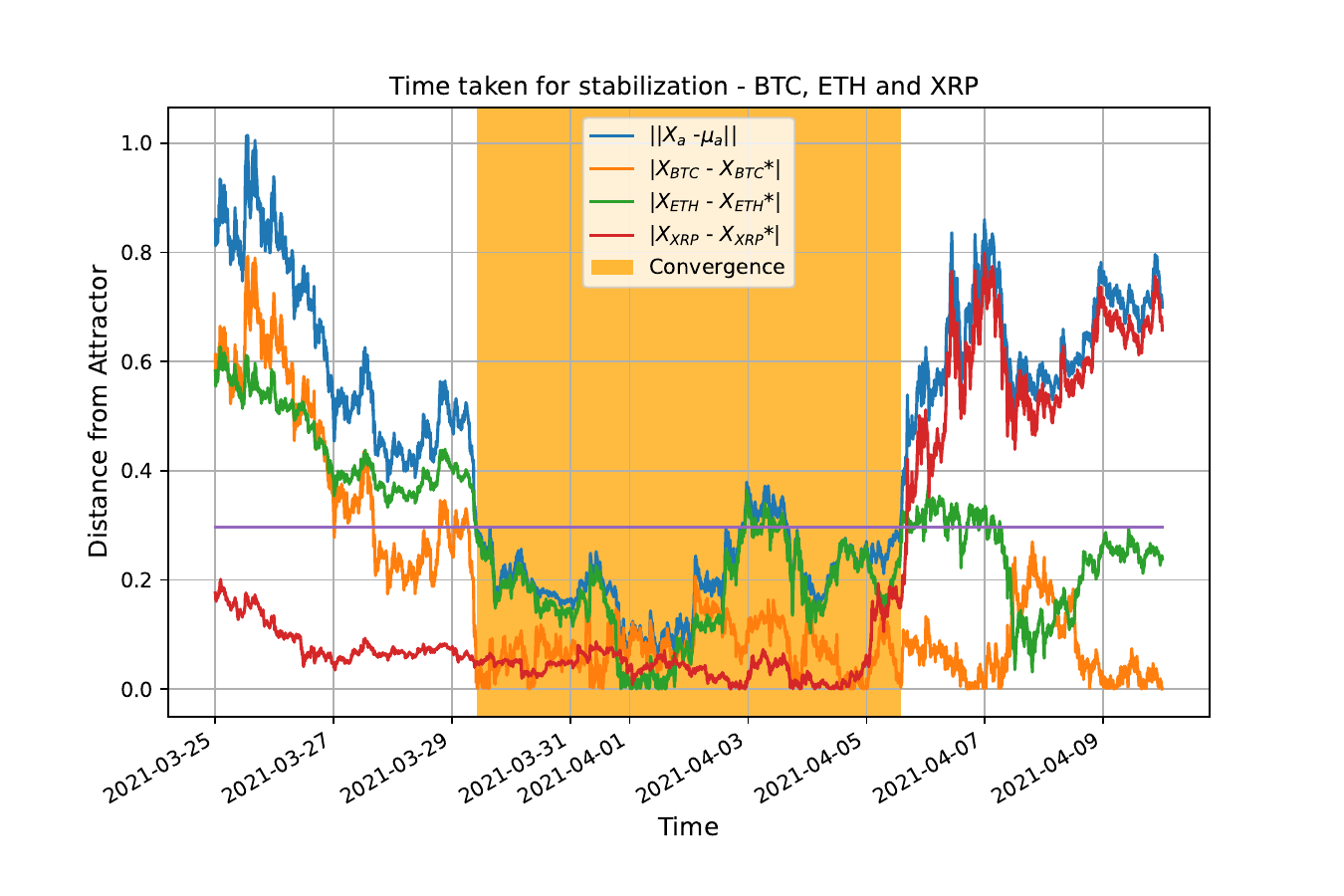}
    \caption{{Convergence to attractor – Before crash - 3D case (25 March to 9 April 2021)}}
    \label{Fig:Conv3dBC}
  \end{minipage}
  \hfill
  \begin{minipage}[t]{0.32\textwidth}
    \includegraphics[width=\textwidth]{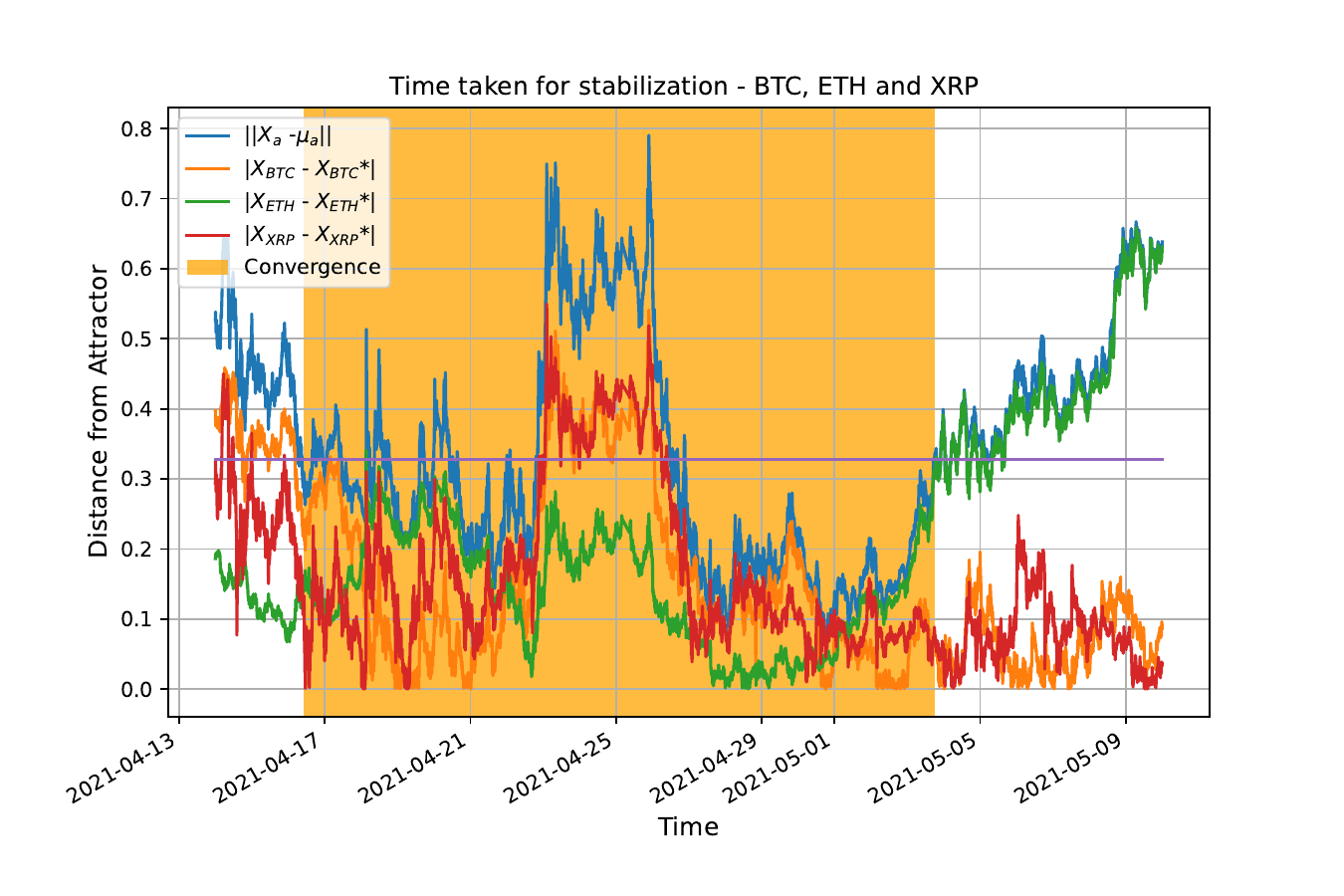}
    \caption{{Convergence to attractor – During crash - 3D case (13 April to 19 May 2021)}}
    \label{Fig:Conv3dDC}
  \end{minipage}
  \hfill
  \begin{minipage}[t]{0.32\textwidth}
    \includegraphics[width=\textwidth]{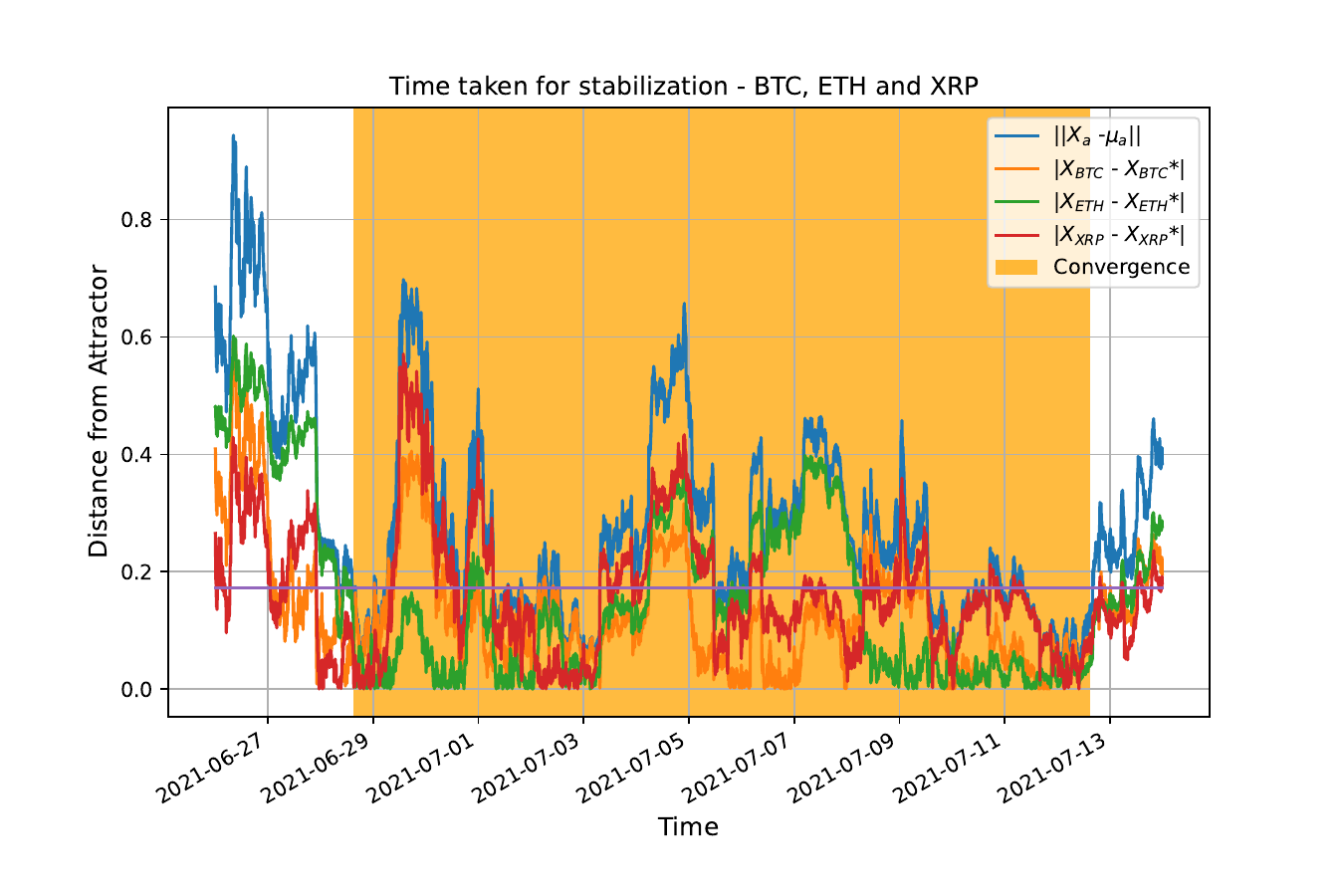}
    \caption{{  Convergence to attractor – After crash - 3D case (26 June to 14 July 2021)}}
    \label{Fig:Conv3dAC}
  \end{minipage}
\end{figure*}

\begin{figure*}[htbp]
  \centering
  \begin{minipage}[t]{0.32\textwidth}
    \includegraphics[width=\textwidth]{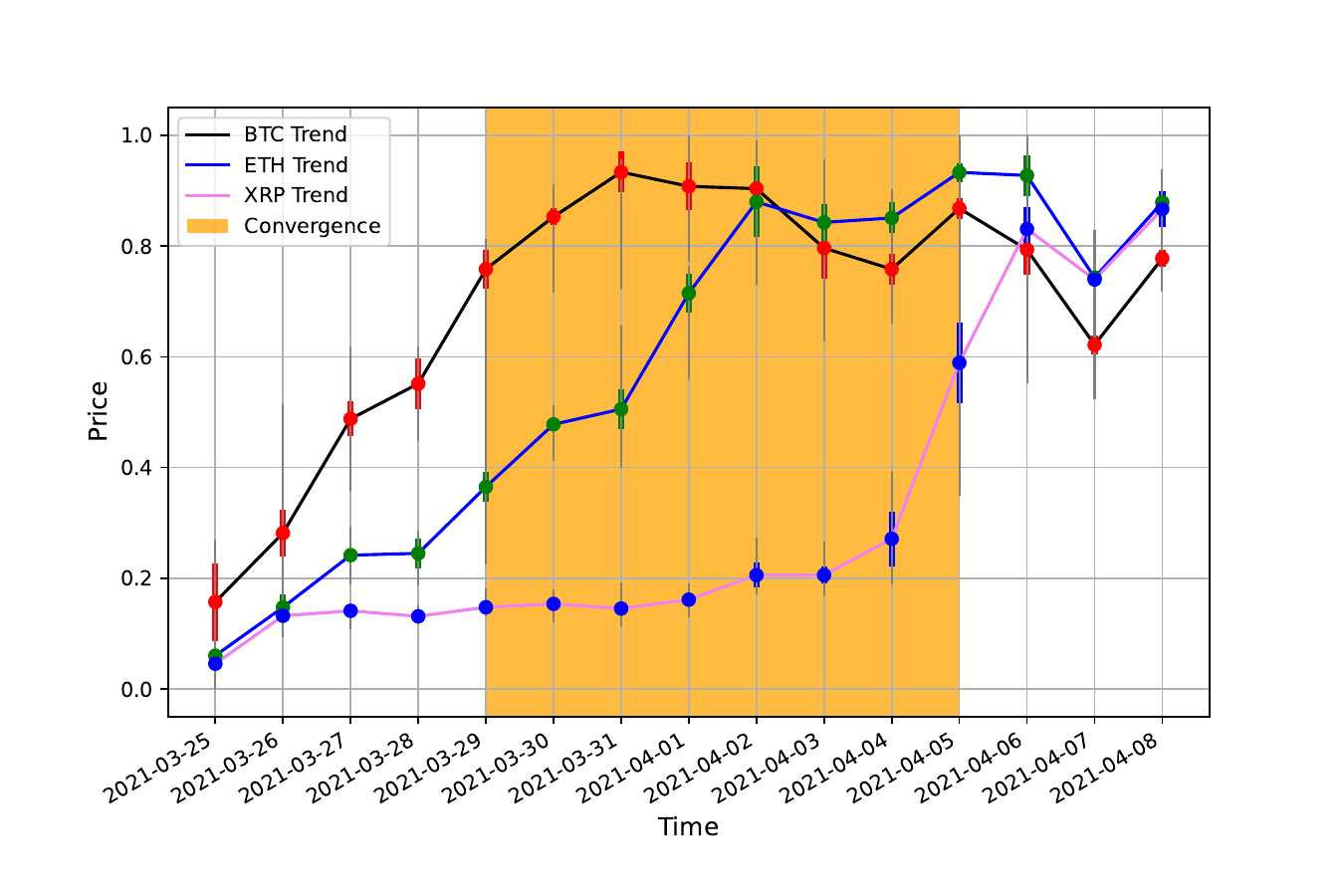}
    \caption{{Temporal evolution of attractor – Before crash} - 3D case (25 March to 9 April 2021)}
    \label{Fig:TrendChangeBC_3D}
  \end{minipage}
  \hfill
  \begin{minipage}[t]{0.32\textwidth}
    \includegraphics[width=\textwidth]{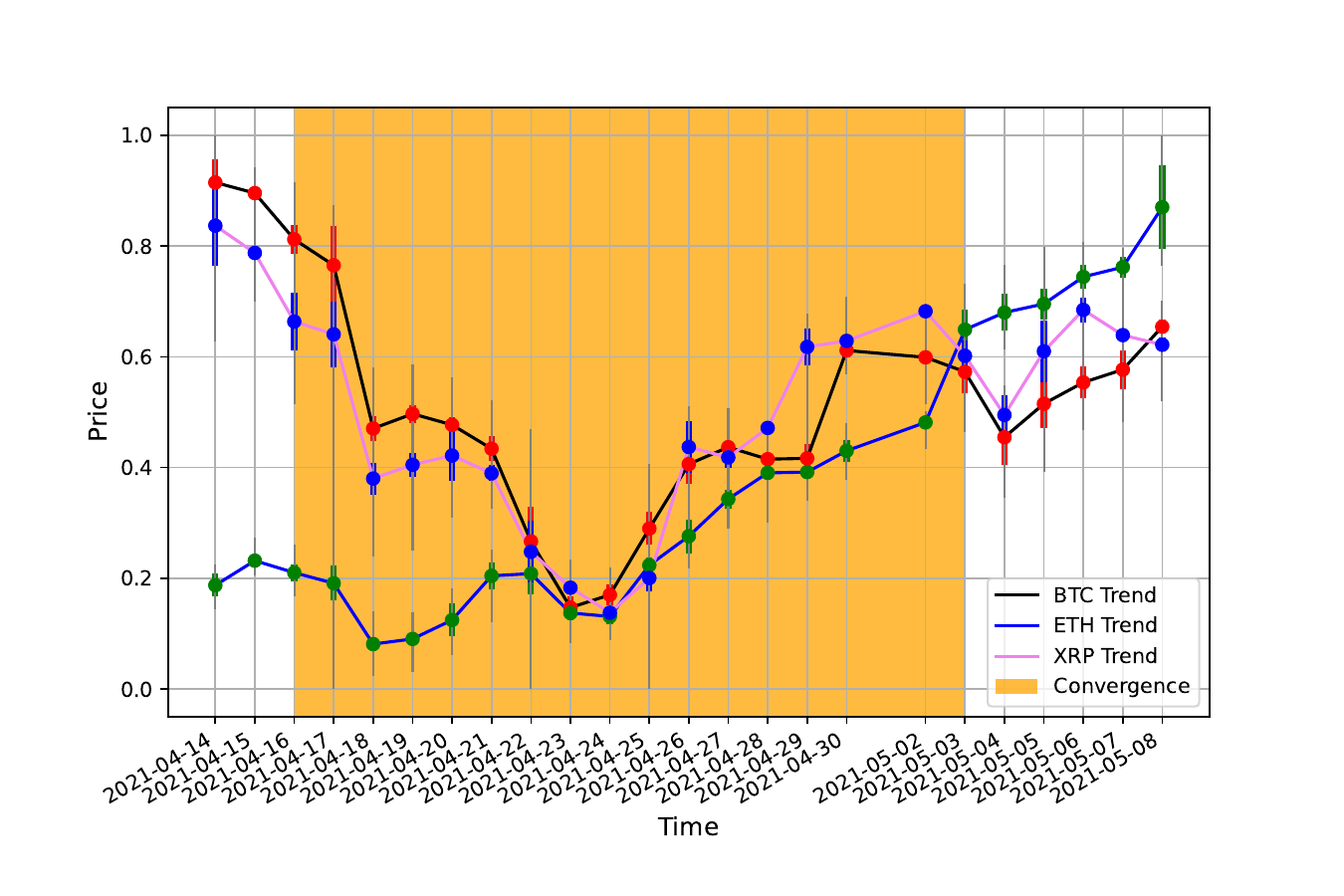}
    \caption{{Temporal evolution of attractor - During crash - 3D case (13 April to 19 May 2021)}}
    \label{Fig:TrendChangeDC_3D}
  \end{minipage}
  \hfill
  \begin{minipage}[t]{0.32\textwidth}
    \includegraphics[width=\textwidth]{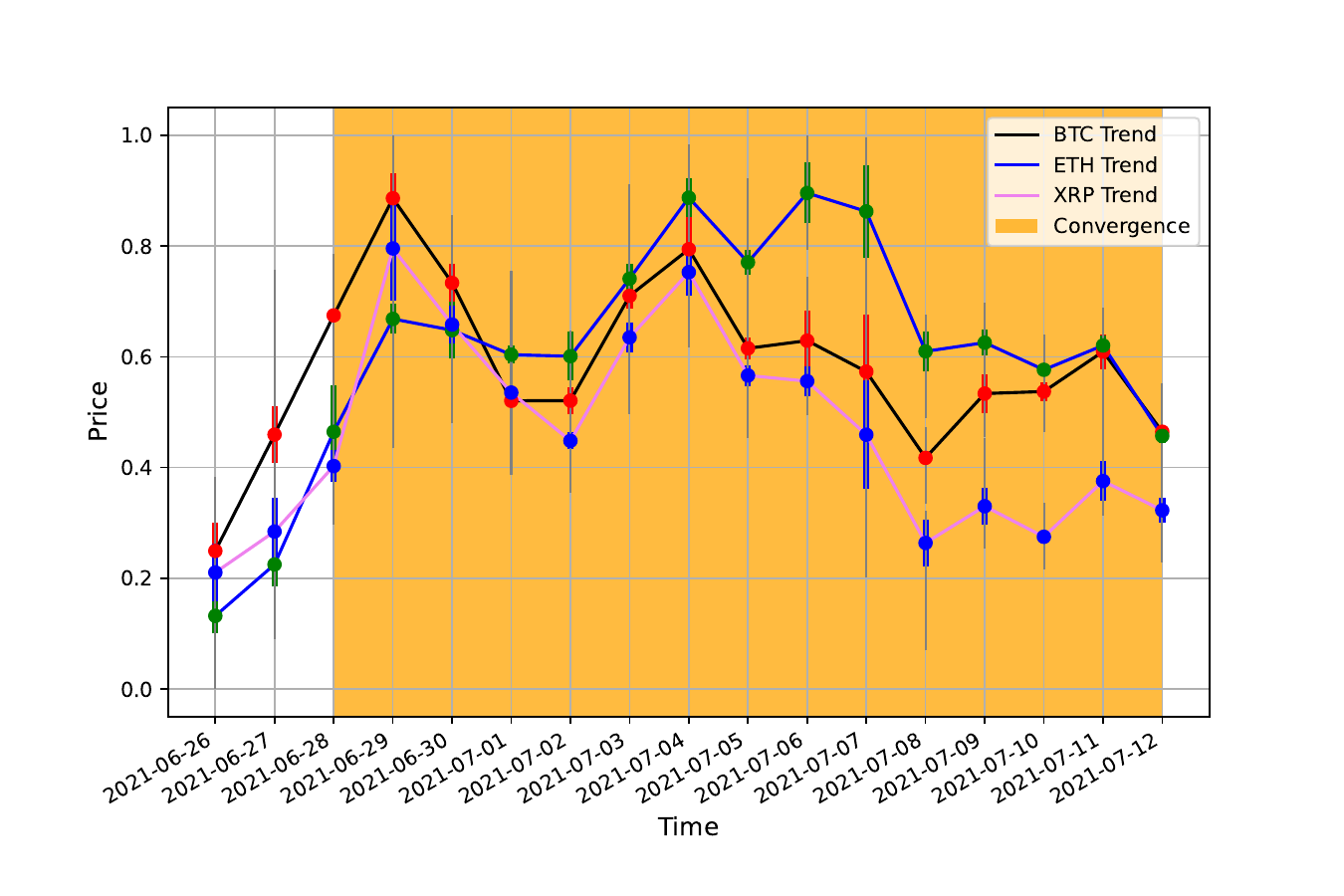}
    \caption{{Temporal evolution of attractor – After crash - 3D case (26 June to 14 July 2021)}}
    \label{Fig:TrendChangeAC_3D}
  \end{minipage}
\end{figure*}

\begin{figure*}[htbp]
  \centering
  \begin{minipage}[t]{0.32\textwidth}
    \includegraphics[width=\textwidth]{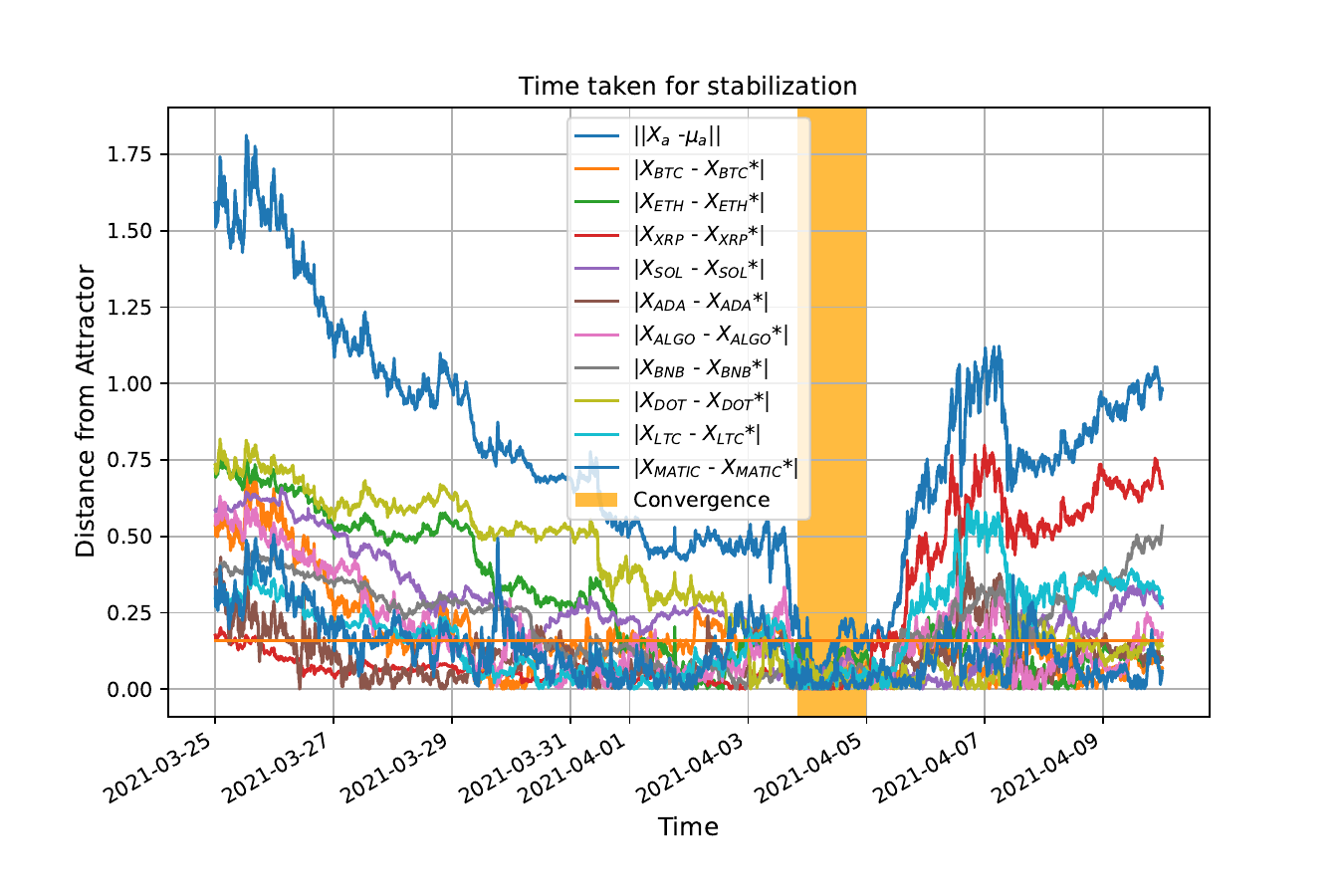}
    \caption{{Convergence to attractor - Before crash - 10D case (25 March to 9 April 2021)}}
    \label{Fig:Conv10dBC}
  \end{minipage}
  \hfill
  \begin{minipage}[t]{0.32\textwidth}
    \includegraphics[width=\textwidth]{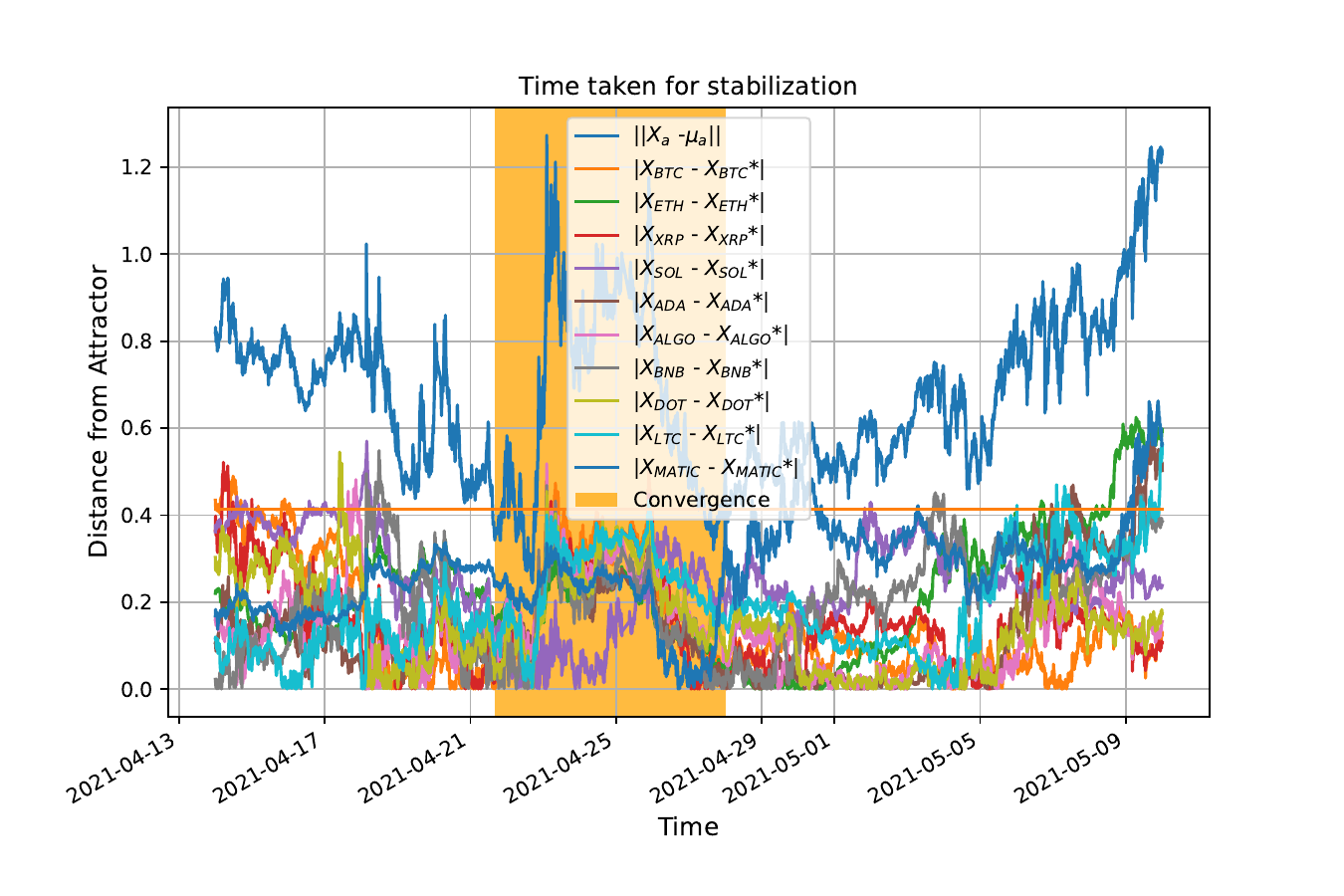}
    \caption{{Convergence to attractor - During crash - 10D case (13 April to 19 May 2021)}}
    \label{Fig:Conv10dDC}
  \end{minipage}
  \hfill
  \begin{minipage}[t]{0.32\textwidth}
    \includegraphics[width=\textwidth]{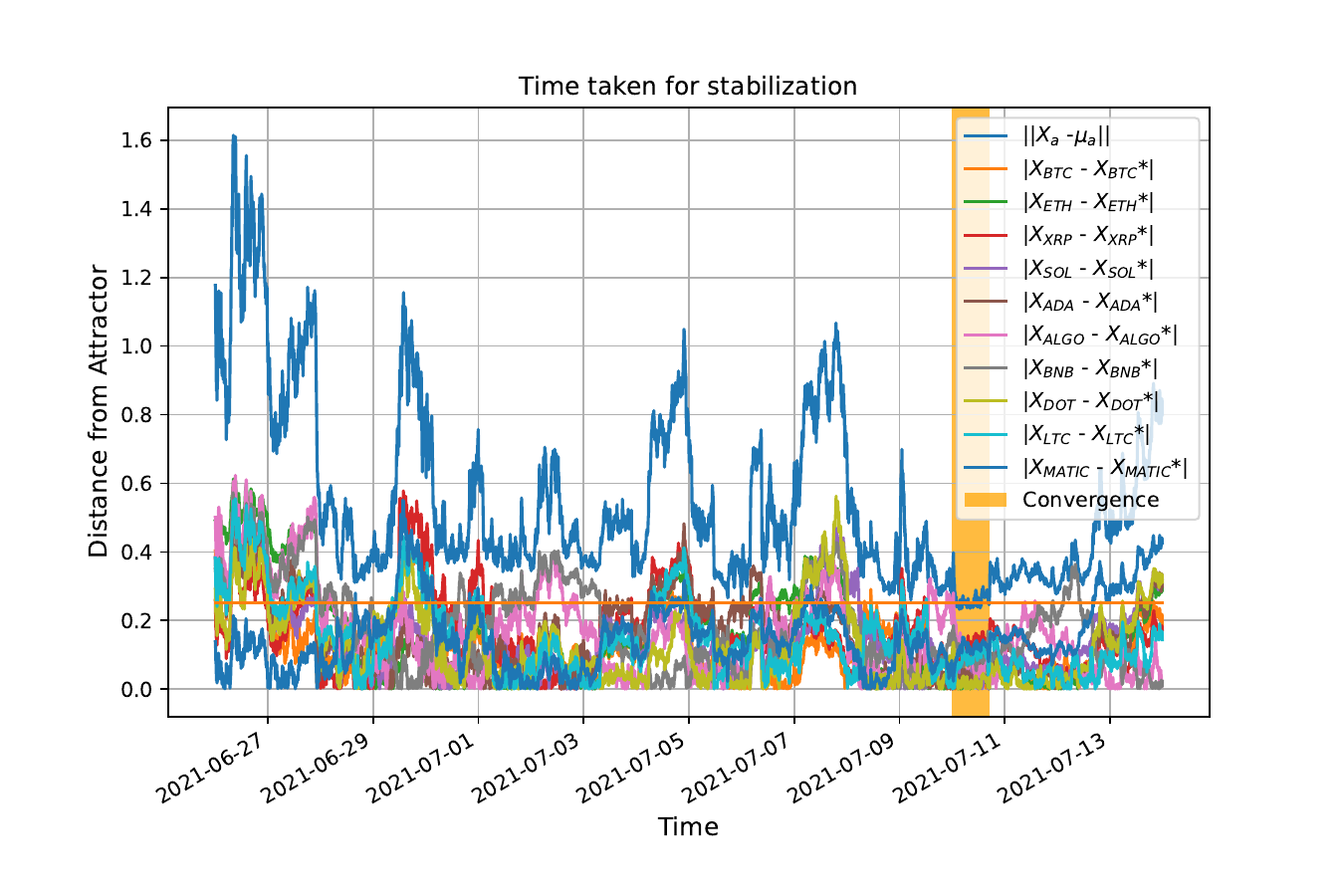}
    \caption{{Convergence to attractor - After crash - 10D case (26 June to 14 July 2021)}}
    \label{Fig:Conv10dAC}
  \end{minipage}
\end{figure*}

\begin{figure*}[htbp]
  \centering
  \begin{minipage}[t]{0.32\textwidth}
    \includegraphics[width=\textwidth]{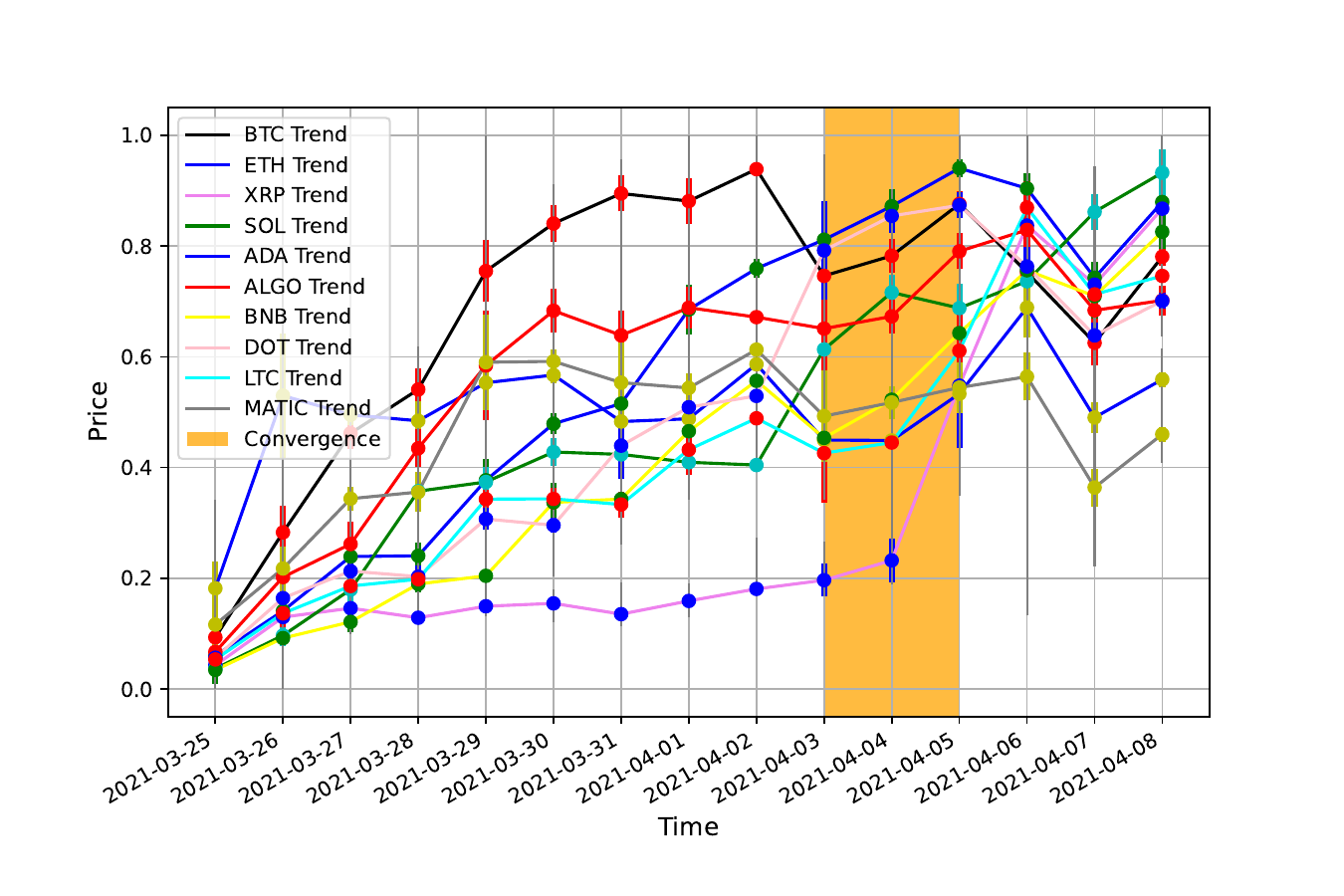}
    \caption{{Temporal evolution of attractor – Before crash - 10D case (25 March to 9 April 2021)}}
    \label{Fig:TrendChangeBC_10D}
  \end{minipage}
  \hfill
  \begin{minipage}[t]{0.32\textwidth}
    \includegraphics[width=\textwidth]{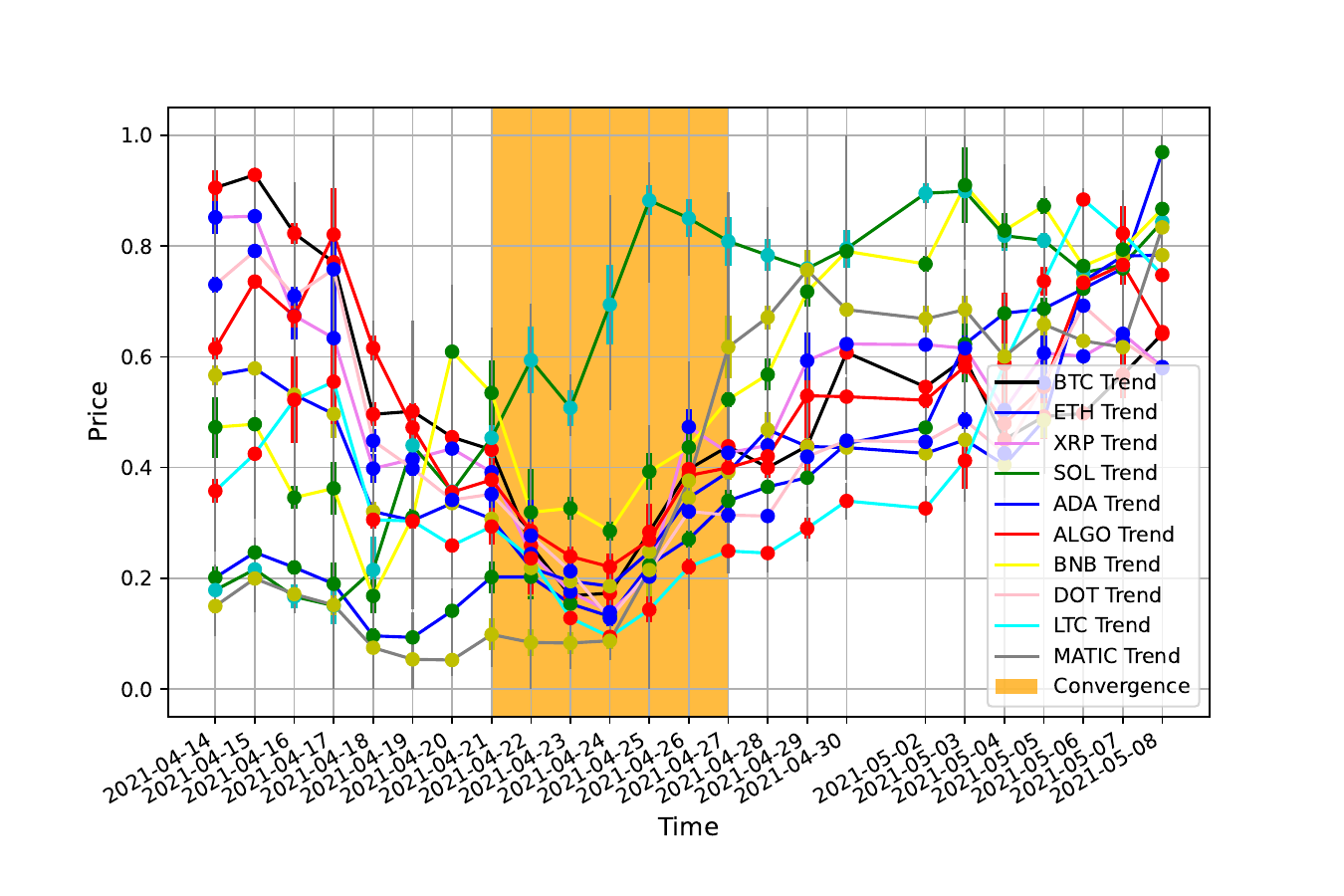}
    \caption{{Temporal evolution of attractor - During crash - 10D case (13 April to 19 May 2021)}}
    \label{Fig:TrendChangeDC_10D}
  \end{minipage}
  \hfill
  \begin{minipage}[t]{0.32\textwidth}
    \includegraphics[width=\textwidth]{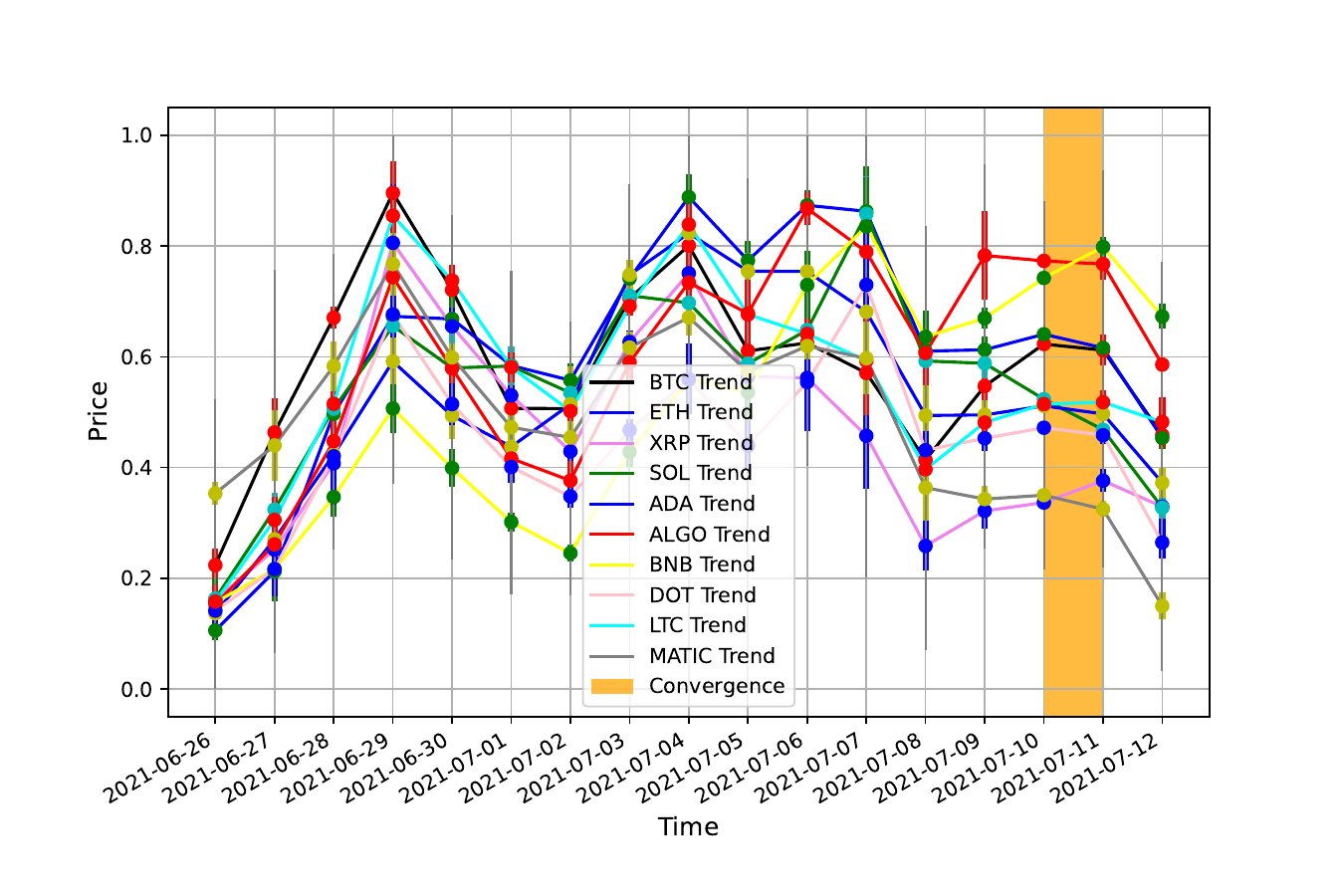}
    \caption{{Temporal evolution of attractor – After crash - 10D case (26 June to 14 July 2021)}}
    \label{Fig:TrendChangeAC_10D}
  \end{minipage}
\end{figure*}

It is relatively easy to extend the analysis to include more than two cryptocurrencies simultaneously. However, representing scenarios with dimensions greater than two using a 2D plot, as done in Figures \ref{BCT2D} through \ref{ACT2D}, is impractical. A more practical approach is to use a facet of figures that illustrate pairwise interactions. To this end, Figures \ref{Fig:Conv3dBC} to \ref{Fig:TrendChangeAC_10D} show the convergence characteristics and temporal evolution of attractors for 3 cryptocurrencies and 10 cryptocurrencies, respectively. Specifically, BTC, ETH, and XRP are considered for the 3 cryptocurrency case, while all 10 cryptocurrencies mentioned in Sec.~{\ref{dataset}} are included in the 10-dimensional case. To aid visualization, the prices of cryptocurrencies in Figures \ref{Fig:Conv3dBC} to \ref{Fig:TrendChangeAC_10D} are normalized using max-min normalization. Whenever distance measures are mentioned, they refer to the distance between the normalized prices.

The results displayed in Figures \ref{Fig:Conv3dBC} to \ref{Fig:TrendChangeAC_10D} are consistent with our 2D analysis inference, as far as BTC and ETH are concerned. The 3D and 10D results notably capture the consistent positive corrections in ETH price across all three windows. Furthermore, they correctly represent the formation of the BTC bubble before the crash, the substantial negative correction of BTC price during the crash, and the recovery of BTC price following the crash. Additionally, Figures \ref{Fig:Conv3dBC} to \ref{Fig:TrendChangeAC_10D} capture the trends and uncertainties of other cryptocurrencies. Other essential parameters, such as market trend, uncertainty, and pairwise correlation, and volatility can be effortlessly quantified for any number of dimensions, and the summary of 3D cases, for all crash durations, are included as a table in the supplementary material.

    \begin{table*}[!htp]\centering
        \caption{Convergence Properties during various observation windows}\label{tab: }
        \scriptsize
        \begin{tabular}{|p{0.13\linewidth}|p{0.11\linewidth}|p{0.11\linewidth}|p{0.11\linewidth}|p{0.13\linewidth}|p{0.12\linewidth}|p{0.07\linewidth}|}
         \hline
         \textbf{Observation window} & \textbf{BTC component of mean attractor} & \textbf{ETH component of mean attractor} & \textbf{XRP component of mean attractor} & \textbf{Onset of Convergence} & \textbf{End of Convergence} & \textbf{Convergence Duration} \\ [0.5ex]
         \hline
            25 Mar to 9 April 2021 &58138.71 &1925.75 &0.59 & Mar 29 2021 & Apr 05 2021 &7 days \\
            \hline
            13 April to 9 May 2021 &56642.38 &2675.17 &1.497 & Apr 16 2021 & May 03 2021 &17 days \\
            \hline
            26 June to 14 July 2021 &33989.65 &2146.28 &0.65 & Jun 28 2021 & Jul 12 2021 &14 days \\
            \hline
            29 to 30 Aug 2021 &48888.295 &3238.814 &1.1533 & Aug 29 2021 & Aug 30 2021 &1 day \\
            \hline
            7 to 9 Sept 2021 &46400.85 &3467.33 &1.101 & Sep 07 2021 & Sep 09 2021 &2 days \\
            \hline
            25 to 26 Sept 2021 &42583.55 &2915.80 &0.941 & Sep 25 2021 & Sep 26 2021 &1 day \\
            \hline
        \end{tabular}
        \label{Tab:Conv3D}
    \end{table*}
\vspace{-1em}

    \begin{table*}[htbp]
        \centering
        \caption{Characterizing trends and uncertainties – BTC, ETH and XRP}
        \begin{tabular}{|p{1.3cm}|p{0.9cm}|p{1cm}|p{1cm}|p{0.8cm}|p{0.9cm}|p{0.9cm}|p{1cm}|p{0.9cm}|p{0.9cm}|p{1cm}|}
            \hline
            \multicolumn{1}{|p{1.3cm}|}{\textbf{Crash Window}} & \multicolumn{1}{p{0.9cm}|}{} &\multicolumn{3}{p{2.8cm}|}{\textbf{Trend Line Magnitude (USD)}} & \multicolumn{3}{p{2.8cm}|}{\textbf{Probability of Negative Correction}} & \multicolumn{3}{p{2.8cm}|}{\textbf{Probability of Positive Correction}}  \\
            \cline{3-5} \cline{6-8} \cline{9-11}
            \multicolumn{1}{|p{1.3cm}|}{} & \multicolumn{1}{p{0.9cm}|}{} & \multicolumn{1}{p{1cm}|}{\textbf{BTC}} & \multicolumn{1}{p{1cm}|}{\textbf{ETH}} & \multicolumn{1}{p{0.8cm}|}{\textbf{XRP}} & \multicolumn{1}{p{0.9cm}|}{\textbf{BTC}} & \multicolumn{1}{p{0.9cm}|}{\textbf{ETH}} & \multicolumn{1}{p{1cm}|}{\textbf{\textbf{XRP}}} & \multicolumn{1}{p{0.9cm}|}{\textbf{BTC}} & \multicolumn{1}{p{0.9cm}|}{\textbf{ETH}} & \multicolumn{1}{p{1cm}|}{\textbf{XRP}} \\
            \hline
            &Before &5750.54 &344.61 &0.11 &0 &0 &0.01 &1 &1 &0.99 \\
            \cline{2-11}
            Apr 2021 &During &-6977.6 &368.76 &-0.31 &1 &0 &1 &0 &1 &0 \\
            \cline{2-11}
            &After &2586.03 &343.65 &0.04 &0 &0 &0.01 &1 &1 &0.99 \\
            \hline
            &Before &627.364 &55.024 &0.0249 &0 &0 &0 &1 &1 &1  \\
            \cline{2-11}
            Sep 2021 &During &-279.2 &56.94 &-0.01 &1 &0 &0.9997 &0 &1 &0.0003 \\
            \cline{2-11}
            &After &-85.92 &-5.45 &0.0004 &1 &1 &0.489 &0 &0 &0.511 \\
            \hline
        \end{tabular}  
        \label{Tab:Trend3D}
    \end{table*}
\vspace{1em}
The convergence characteristics, trend characterization and structural dependence during convergence of BTC-ETH-XRP case are summarized in Tables \ref{Tab:Conv3D}, \ref{Tab:Trend3D} and \ref{Tab:Struct3D}, respectively. It can be observed in comparison with 2-dimensional case (BTC-ETH) that the duration of convergence has substantially reduced in the three-dimensional case. For example, during April 2021 crash window, a crash that has sustained for a long period, the duration of convergence were 20 and 8 for 2-dimensional and 3-dimensional cases respectively. This is a consequence of very high volatility of XRP. Similarly, the probability of positive or negative corrections are almost certain for BTC and ETH, while there is a significant uncertainty in the case of XRP (Table \ref{Tab:Trend3D}). Further, Table~\ref{Tab:Struct3D} quantifies the structural dependence and volatility between BTC, ETH and XRP during convergence.  

    \begin{table*}[!htp]\centering
        \caption{Structural dependence during convergence – BTC, ETH and XRP}\label{tab: }
        \scriptsize
        \begin{tabular}{|p{0.13\linewidth}|p{0.07\linewidth}|p{0.13\linewidth}|p{0.09\linewidth}|p{0.09\linewidth}|p{0.09\linewidth}|}
         \hline
         \textbf{Observation window} & \textbf{Eigen Value} & \textbf{Eigen Vector} & \textbf{Phase (BTC - ETH)} & \textbf{Phase (ETH - XRP)} & \textbf{Phase (BTC - XRP)} \\ [0.5ex]
         \hline
            25 Mar to 9 April 2021 &0.036 &[0.080, 0.758, 0.646]  &83.93\degree &40.45\degree &7.04\degree \\
            \hline
            13 April to 9 May 2021  &0.044 &[-0.814, 0.362, -0.453] &156.01\degree &129\degree &27.5\degree \\
            \hline
            26 June to 14 July 2021  &0.012 &[-0.299,  0.947,  0.109] &107.5\degree &11.51\degree &24\degree \\
            \hline
            29 to 30 Aug 2021 &0.001 &[0.088, 0.996, 0.00] &84.9\degree &0.00\degree &0.00\degree \\
            \hline
            7 to 9 Sept 2021 &0.008 &[-0.466, 0.727, 0.503] &122.67\degree &34.66\degree &132.8\degree \\
            \hline
            25 to 26 Sept 2021 &0.015 &[-0.176,  0.0006,  0.984] &0.00\degree &90\degree &45.6\degree \\
            \hline
        \end{tabular}
        \label{Tab:Struct3D} 
    \end{table*}
\subsection{Comparison with wavelet coherence technique}\label{WaveletCompar}

In this section, we analyse the consistency of the structural dependence information inferred using the potential field approach, with the results obtained using the popular wavelet coherence based technique for analysing structural dependence between any given pairs of cryptocurrencies \cite{existingWavelet}. Wavelet coherence technique analyse the dependency between two time series, in the time-frequency domain, by extending the concept of cross-correlation to the wavelet transform framework. 

The cross wavelet transform of two time series $x(t)$ and $y(t)$ is given by
\begin{equation}
    W_{xy}(a, b) = \frac{1}{a} \int_{-\infty}^{\infty} x(t) \cdot y^*\left(\frac{t-b}{a}\right) \cdot \psi^*\left(\frac{t-b}{a}\right) dt,
\end{equation}
where, $W_{xy}(a, b)$ represents the cross wavelet coefficients as a function of scale (frequency) $a$ and translation (time) $b$.
As in \cite{existingWavelet}, we use a Morlet wavelet, given by \eqref{Morlet},
\begin{equation}\label{Morlet}
    \psi(t) = \pi^{-1/4} e^{i \omega_0 t} e^{-t^2/2},
\end{equation}
where, $\omega_0$ is the central frequency of the wavelet.
Further, wavelet coherence~$R_{xy}$ between the two time series $x(t)$ and $y(t)$ can be computed as,
\begin{equation}\label{wavecoh}
    R^2_{xy} = \frac{|S(a^{-1}W_{xy}(a, b))|^2}{S(a^{-1}|W_{x}(a, b)|^2)S(a^{-1}|W_{y}(a, b)|^2)}.
\end{equation}
In \eqref{wavecoh}, $S(\dot)$ is the smoothing function, for example, weighted running average in both time and frequency directions. Further, $W_{x}(a, b)$ and $W_{y}(a, b)$ are continuos wavelet transforms of $x(t)$ and $y(t)$, respectively. It is easy to see that $0 \leq R^2_{xy} \leq 1$. When the value of $R^2_{xy}$ is close to zero, it indicates that the correlation between the time series is weak, whereas, a value of $R^2_{xy}$ close to $1$ signifies a strong correlation. The coherence phase, $tan^{-1}\left(\mathbb{R}e(W_{xy}(a,b))/\mathbb{I}m(W_{xy}(a,b))\right)$, represents the phase angle associated with the coherence between the two time series at different time-frequency points. It gives information about the time lag or phase relationship between the two time series. 
    \begin{figure*}
      \centering
      \begin{minipage}[t]{0.32\textwidth}
        \includegraphics[width=\textwidth]{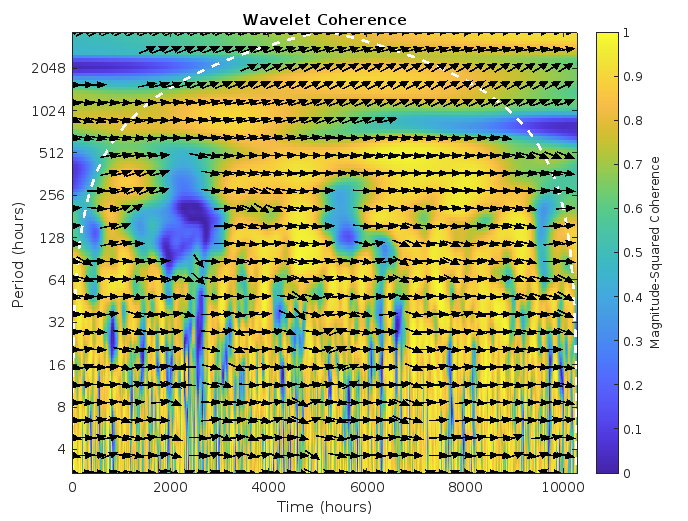}
        \caption{{Wavelet coherence between BTC and ETH prices before crash period (25 March to 9 April 2021)}}
        \label{WPlotBC}
      \end{minipage}
      \hfill
      \begin{minipage}[t]{0.32\textwidth}
        \includegraphics[width=\textwidth]{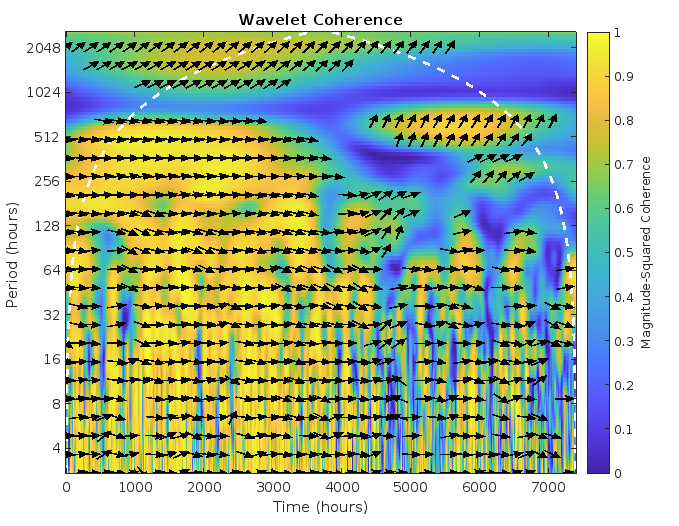}
        \caption{{Wavelet coherence between BTC and ETH prices during crash period (13 April to 19 May 2021)}}
        \label{WPlotDC}
      \end{minipage}
      \hfill
      \begin{minipage}[t]{0.32\textwidth}
        \includegraphics[width=\textwidth]{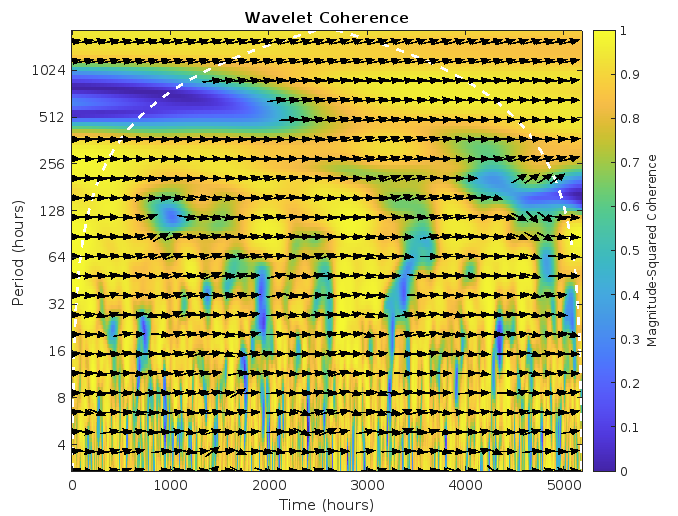}
        \caption{{Wavelet coherence between BTC and ETH prices after crash period (26 June to 14 July 2021)}}
        \label{WPlotAC}
      \end{minipage}
    \end{figure*}

The wavelet coherence plots, Figures \ref{WPlotBC} through \ref{WPlotAC}, show the coherence, $R_{xy}^2$ (computed using \eqref{wavecoh}), between the time series of BTC and ETH prices, as a function of frequency and time, along with the coherence phase. The horizontal axis of the wavelet coherence plots represent the timescale, and the vertical axis represent the time period (inverse of frequency) scale of coherence. The strength of correlation is indicated by the colour map, with yellow indicating, $R_{xy}^2=1$, and blue indicating, $R_{xy}^2 = 0$. For each frequency and time combination, the coherence phase determines the direction of arrows in the wavelet coherence plots. A positive correlation between the two time series is represented by an arrow directed towards the right, while a negative correlation is denoted by an arrow pointing towards the left. Additionally, the orientation of the arrow (upward or downward) provides information about the temporal relationship between the two signals. Specifically, if the arrow points upwards, it indicates that the first time series, BTC price, leads the second time series, ETH price, at that particular time-frequency point. Conversely, if the arrow points downwards, it signifies that the ETH price leads the BTC price at that specific time and frequency. Arrows present in the lower time period horizon indicate short term dependency, whereas arrows in the upper time period horizon shows long term dependency, between BTC and ETH price variations.

The inferences regarding structural dependence, between BTC and ETH, drawn from Figures \ref{WPlotBC} through \ref{WPlotAC} can be summarized as follows: \begin{inparaenum}[(i)]
    \item Before the crash, from 25 March to 9 April 2021, the wavelet coherence analysis (Fig. \ref{WPlotBC}) reveals predominant yellow regions and right-directed arrows, indicating a positive correlation between BTC and ETH prices.
    \item During the crash period, from 13 April to 19 May 2021, significant blue regions appear (Fig. \ref{WPlotDC}), suggesting a very low correlation between BTC and ETH prices. Moreover, the presence of left-up arrows during the crash period indicates a negative correlation. This negative correlation pattern signifies that when BTC prices dropped, there was a subsequent gain in ETH prices, suggesting a shift of assets from BTC to ETH.
    \item After the crash (Fig. \ref{WPlotAC}), from 26 June to 14 July 2021, the wavelet coherence analysis shows an increased yellow area compared to the other two cases, along with a rightward arrow. This indicates a strengthened correlation between BTC and ETH prices during the post-crash duration. In summary, the structural dependence between BTC and ETH has heightened after the crash event.
\end{inparaenum}

The aforementioned inferences obtained from the popular wavelet coherence approach are consistent with the inferences on structural dependence, as in Table~{\ref{2DStructDep}}, made using the proposed potential field approach (Sec. \ref{StrDepPot}). 

\subsection{Improvement in Error Performance of LSTM based Prediction Techniques when Mean Attractor is used as an input feature}\label{LSTMComp}
\begin{figure}[htbp]
\vspace{-1em}
  \centering
    \includegraphics[width=0.6\linewidth]{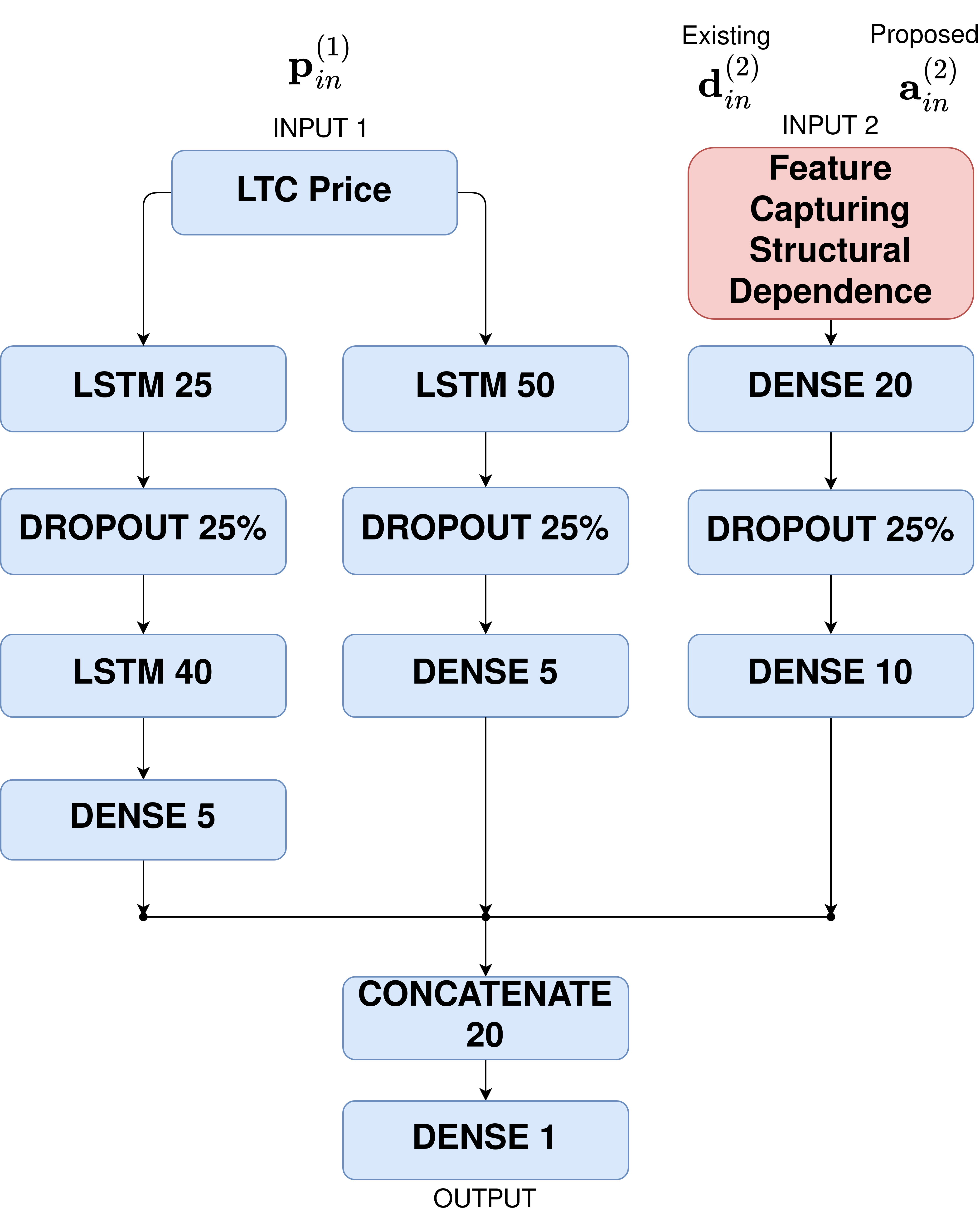}
    \caption{{  LSTM based price prediction of LTC – The existing model \cite{existingLSTM} uses the BTC price direction, $\mathbf{d}_{in}^{(2)}(i)$, as the input 2, the feature capturing structural dependence. The proposed approach substitutes the input 2 in the existing approach with the BTC component of the posterior mean of mean attractor, $\mathbf{a}_{in}^{(2)}(i)$, for better error performance.}}
    \label{LSTMArch}    
\end{figure}

Many deep learning models are available in literature that predict cryptocurrency price based on historic price information.
\cite{existingLSTM} presented an LSTM based model which exploits the interdependence between LTC and BTC to improve the performance, in terms of MSE and MAE of LTC price prediction, compared to the LSTM models that considers historic LTC prices alone as input features. The architecture used in \cite{existingLSTM} is shown in Fig.~\ref{LSTMArch}. This LSTM model takes in two inputs and generates the LTC price as output. The first input is a historical record of LTC's daily closing prices, $\zeta_{1:N} = \{p_1, p_2, \cdots, p_N\}$, while the second input is a historical record of some attribute, $\theta_{1:N} = \{\tau_1, \tau_2, \cdots, \tau_N \}$, that can capture the daily market trend of BTC, the parent coin. In \cite{existingLSTM}, the attribute used for the second input is the BTC price direction, $\theta_{1:N}^{dir} = \{\tau_1^{dir}, \tau_2^{dir}, \cdots, \tau_N^{dir} \}$, which is determined by comparing the daily mean price of BTC to its open price. If the mean price is greater than the open price, the corresponding BTC price direction attribute, $\tau_i^{dir}$, is set to 1; otherwise, it is set to -1.

However, the BTC price direction attribute used in \cite{existingLSTM}, depends on the mean price, of a time series, that exhibits fluctuations, due to significant volatility in the market, and hence, in this paper we argue that the mean attractor inferred from the time series is a more reliable indicator of the market trend and structural dependence than the BTC price direction attribute. We propose, using the mean attractor of the daily price window (computed as in \eqref{weigtedMu}), $\theta_{1:N}^{att} = \{\tau_1^{att}, \tau_2^{att}, \cdots, \tau_N^{att} \}$, instead the BTC price direction attribute, $\theta_{1:N}^{dir}$, as the second input to the model in \cite{existingLSTM}. 

To compare the error performance of the existing and proposed approaches, we used historical data of BTC and LTC closing prices sampled every 5 minutes from 1 June 2020 to 31 May 2021. All training parameters were kept constant except for input. For input 1, the daily closing prices of LTC were calculated. We computed the BTC component of the posterior mean of the mean attractor of each one day window (considering BTC and LTC), and the BTC price direction, for input 2 of proposed and existing approaches, respectively. To perform LSTM regression, we used 5 time lags (the previous 5 prices) of the daily closing prices of LTC as input 1 for both existing and proposed approaches as in \eqref{input1}. 
\begin{equation}\label{input1}
    \mathbf{p}_{in}^{(1)}(i) = \left[p_i, p_{i-1},\cdots,p_{i-4}\right].
\end{equation}
Whereas, the input 2 were, 5 time lags of  the mean attractors and the parent direction of BTC, for proposed and existing approaches, respectively as in \eqref{input2att} and \eqref{input2dir}. 
\begin{equation}\label{input2dir}
    \mathbf{d}_{in}^{(2)}(i) = \left[\tau_i^{dir}, \tau_{i-1}^{dir},\cdots,\tau_{i-4}^{dir}\right],
\end{equation}and 
\begin{equation}\label{input2att}
    \mathbf{a}_{in}^{(2)}(i) =\left[\tau_i^{att}, \tau_{i-1}^{att},\cdots,\tau_{i-4}^{att}\right].
\end{equation}
The corresponding output in both the approaches is,
\begin{equation}\label{output}
    \mathbf{o}(i) = \left[p_{i+1}\right].
\end{equation}In \eqref{input1} through \eqref{output}, $i = 5,\cdots,N-1$, where $N$ is the total number of training samples.
\vspace{-1em}
    
    \begin{table}[!htp]\centering
        \caption{LSTM based price prediction – MSE performance comparison}
     \scriptsize
       \scalebox{0.9}{ \begin{tabular}
        {|p{1cm}|p{0.5cm}|p{0.5cm}|p{0.5cm}|p{0.5cm}|p{0.5cm}|p{0.5cm}|p{0.5cm}|p{0.5cm}|p{0.5cm}|p{0.5cm}|p{0.5cm}|}
            \hline
            &\textbf{1 Day} &\textbf{2 Day} &\textbf{3 Day} &\textbf{5 Day} &\textbf{7 Day} &\textbf{10 Day} &\textbf{12 Day} &\textbf{13 Day} &\textbf{14 Day}  \\
            \hline
            \textbf{Mean Existing} &0.073 &0.081 &0.089 &0.106 &0.139 &0.173 &0.201 &\textbf{0.213} &\textbf{0.225}  \\
            \hline
            \textbf{STD Existing} &0.147 &0.179 &0.207 &0.27 &0.355 &0.435 &0.471 &0.487 &0.499  \\
            \hline
            \textbf{Mean Proposed} &\textbf{0.059} &\textbf{0.067} &\textbf{0.074} &\textbf{0.090} &\textbf{0.107} &\textbf{0.143} &\textbf{0.159} &0.224 &0.266  \\
            \hline
            \textbf{STD Proposed} &0.139 &0.17 &0.202 &0.266 &0.347 &0.486 &0.53 &1.259 &1.791  \\
            \hline
        \end{tabular}}
        \label{MSEComp} 
    \end{table}
\vspace{-1em}
    \begin{table}[!htp]\centering
        \caption{LSTM based price prediction – MAE performance comparison}
        \scriptsize
        \scalebox{0.9}{\begin{tabular}
        {|p{1cm}|p{0.5cm}|p{0.5cm}|p{0.5cm}|p{0.5cm}|p{0.5cm}|p{0.5cm}|p{0.5cm}|p{0.5cm}|p{0.5cm}|p{0.5cm}|p{0.5cm}|}
            \hline
            &\textbf{1 Day} &\textbf{2 Day} &\textbf{3 Day} &\textbf{5 Day} &\textbf{7 Day} &\textbf{10 Day} &\textbf{12 Day} &\textbf{13 Day} &\textbf{14 Day}  \\
            \hline
            \textbf{Mean Existing} &0.198 &0.204 &0.209 &0.219 &0.253 &0.284 &0.311 &0.322 &0.333  \\
            \hline
            \textbf{STD Existing} &0.183 &0.190 &0.196 &0.211 &0.238 &0.260 &0.273 &0.278 &0.283  \\
            \hline
            \textbf{Mean Proposed} &\textbf{0.176} &\textbf{0.182} &\textbf{0.188} &\textbf{0.197} &\textbf{0.210} &\textbf{0.231} &\textbf{0.244} &0.254 &0.267  \\
        \hline
        \textbf{STD Proposed} &0.168 &0.173 &0.182 &0.196 &0.216 &0.284 &0.259 &0.288 &0.313  \\
        \hline
        \end{tabular}}
        \label{MAEComp} 
    \end{table}

\vspace{2em}
For the existing approach, the model was trained using the input, $[\mathbf{p}_{in}^1(i),\mathbf{d}_{in}^2(i)]$, and, output $\mathbf{o}(i)$, where $i = 5,\cdots,N-1$. Similarly, for the proposed approach, the model was trained using the input $[\mathbf{p}_{in}^1(i),\mathbf{a}_{in}^2(i)]$, and, output $\mathbf{o}(i)$, where $i = 5,\cdots,N-1$. The estimated price of day $N+1$, namely $\hat{p}_{N+1}$, were predicted using the respective trained models, with $[\mathbf{p}_{in}^1(N),\mathbf{d}_{in}^2(N)]$ and $[\mathbf{p}_{in}^1(N),\mathbf{a}_{in}^2(N)]$, as the test inputs for the existing and proposed approaches, respectively. Further, we obtained $\hat{\mathbf{p}}_{in}^1(N+1) = [\hat{p}_{N+1}, p_N, p_{N-1},p_{N-2}, p_{N-3}]$, the estimate of $\mathbf{p}_{in}^1(N+1)$, making use of the latest prediction, $\hat{p}_{N+1}$. Subsequently, $\hat{p}_{N+2}$, the estimate of $p_{N+2}$ were obtained using $\hat{\mathbf{p}}_{in}^1(N+1)$ as test input 1 to the respective models. This process was repeated to obtain predictions over time horizons, up to $14$ days.

The performance metric, MSE and MAE over $T$ day prediction is computed as in \eqref{MSEEq} and \ref{MAEEq} respectively.
\begin{equation}\label{MSEEq}
    MSE = \frac{1}{T}\sum_{i = N+1}^{N+T}\left(p_i - \hat{p}_i\right)^2, \text{ }T = 1,2,\cdots,24.
\end{equation}

\begin{equation}\label{MAEEq}
    MAE = \frac{1}{T}\sum_{i = N+1}^{N+T}\left|p_i - \hat{p}_i\right|, \text{ }T = 1,2,\cdots,24
\end{equation}

In order to compare the prediction performances in terms of MSE and MAE, we predicted LTC prices up to a time horizon of 24 days for various values of $N$, ranging from 90 to 340, using a batch size of 10, over 1000 epochs, and the MSE and MAE values were observed for all iterations. The comparison of MSE and MAE performances up to a time horizon of 14 days are given in Tables \ref{MSEComp} and \ref{MAEComp}, respectively. It can be observed from the Table~{\ref{MSEComp}} that the mean MSE of 1 day prediction of the existing method is 0.07, however, it has reduced by almost $28\%$ when the mean attractor is used as one of the input attributes, to 0.05. It can also be observed from the Table~{\ref{MSEComp}} that, the mean attractor based approach consistently performs better in terms of mean MSE over all the time horizons of predictions, till 12 days. The mean MSE of 12 day prediction of the proposed approach is $25\%$ better compared to the existing approach. A better MSE performance of the proposed approach signifies that the use of mean attractor as one of the input features, improves the ability of the predictive model to capture extreme events in LTC price variations. However, after 12 days, the average MSE of the proposed method degrades due to the accumulation of uncertainty contributed by the Gaussian Process used to estimate the mean attractor, resulting in a significant number of extreme outliers. However, the MSE performance of the proposed approach, up to the 75th percentile remains consistently better (on an average by $55\%$) for all prediction horizons, both above and below the 12-day mark, when compared to the existing method. The detailed numerical results of MSE and MAE performances of prediction, up to a time horizon of 24 days, are included in the supplementary material.

\vspace{1em}
The MAE performance comparison of both existing and the proposed approaches is given in Table \ref{MAEComp}. As can be seen from the numerical results, the proposed approach performs consistently better in terms of average MAE, for all the time horizons of predictions, despite the significantly extreme outliers, attributed by the uncertainty propagation. This implies that the use of mean attractor as one of the input attributes have improved the average performance of the LSTM based predictive model.

\vspace{1em}
\section{Conclusion}\label{Concl}
Identifying the structural dependence, volatility, and the trends of future price corrections are fundamental to cryptocurrency trading.
Owing to extreme volatility in cryptocurrency prices, characterizing the market dynamics, and inferring meaningful features that can capture the price correction trends and structural dependence, based on historic price information is challenging, and it in turn affects, the performance of state-of-the-art machine learning models used in algorithmic trading applications. This paper used a Bayesian Machine Learning technique, namely, Gaussian Process, and the Potential Field theory to develop a unified framework that can simultaneously characterize and quantify: \begin{inparaenum}[(a)] \item trends in price movement of cryptocurrencies, \item structural dependencies between various cryptocurrencies, and \item volatility. \end{inparaenum}  Using the proposed potential field approach, we have analysed the cryptocurrency market data during different BTC crash durations and \begin{inparaenum}[(i)]
    \item inferred mean attractors and characterized the volatility, and, uncertainty in price corrections of the cryptocurrencies, \item characterized the structural dependencies between various cryptocurrencies, \item inferred the direction of asset movement between various cryptocurrencies, \item analysed the temporal evolution 
 of attractors and convergence characteristics of the cryptocurrency market, and \item visualized the structural dependencies, price correction trends and associated uncertainties with the help of self-explanatory diagrams. 
\end{inparaenum} A comparison with the wavelet coherence based approach showed that the structural dependency information inferred by the proposed potential field based approach is consistent with the results of popular wavelet coherence approach, while generalizing the approach to more than $2$ cryptocurrencies. Further, the inferred mean attractor, when used as an input feature in LSTM based techniques, on an average, gave $25\%$ better MSE and MAE performance compared to the existing techniques, in predicting the future price of cryptocurrencies.

\section*{Acknowledgement}
This work was partly supported by the Department of Science and Technology, Government of India through the FIST Scheme under the Grant SR/FST/ET-I/2017/68. 

\bibliographystyle{ieeetr}
{\small \bibliography{zpaper_Ref}}

\end{document}